\documentclass[conference]{IEEEtran}
\IEEEoverridecommandlockouts
\usepackage{cite}
\usepackage{amsmath,amssymb,amsfonts}
\usepackage{algorithmic}
\usepackage{algorithm}
\usepackage{booktabs}
\usepackage{listings}
\usepackage{multirow}
\usepackage[normalem]{ulem}
\useunder{\uline}{\ul}{}
\makeatletter
\newcommand{\removelatexerror}{\let\@latex@error\@gobble}
\makeatother
\usepackage{graphicx}
\usepackage{textcomp}
\usepackage{xcolor}
\def\BibTeX{{\rm B\kern-.05em{\sc i\kern-.025em b}\kern-.08em
    T\kern-.1667em\lower.7ex\hbox{E}\kern-.125emX}}
    
\usepackage{xspace}

\newif\ifsubmission
\submissionfalse

\ifsubmission

\else

\fi

\newcommand{\sys}{CIAO\xspace}
\newcommand{\stitle}[1]{\vspace{1ex} \noindent{\bf #1}}
    
\begin{document}

\title{\sys: An Optimization Framework for Client-Assisted Data Loading}

\author{\IEEEauthorblockN{Cong Ding\textsuperscript{\textsection}}
\IEEEauthorblockA{\textit{Peking University}\\
congding@pku.edu.cn}
\and
\IEEEauthorblockN{Dixin Tang, Xi Liang, Aaron J. Elmore, Sanjay Krishnan}
\IEEEauthorblockA{\textit{The University of Chicago}\\
\{totemtang@, xiliang@, aelmore@cs., skr@\}uchicago.edu}
}


\maketitle
\thispagestyle{plain}
\pagestyle{plain}

\begingroup\renewcommand\thefootnote{\textsection}
\footnotetext{Work done at ChiData group of the University of Chicago.}
\endgroup

\begin{abstract}
Data loading has been one of the most common performance bottlenecks for many big data applications, especially when they are running on inefficient human-readable  formats, such as JSON or CSV.
Parsing, validating, integrity checking, and data structure maintenance are all computationally expensive steps in loading these formats. 
Regardless of these costs, many records may be filtered later during query evaluation due to highly selective predicates -- resulting in wasted computation. 
Meanwhile, computing power of client ends are typically not exploited.
Here, we explore investing limited cycles of clients on prefiltering to accelerate data loading and enable data skipping for query execution.
In this paper, we present \sys, a tunable system to enable client cooperation with the server to enable efficient partial loading and data skipping for a given workload.
We proposed an efficient algorithm which would select near-optimal predicate set to push down within given budget.
Moreover, \sys will address the trade-off between client cost and server savings by setting different budgets for different clients.
We implemented \sys and evaluated its performance on three real-world datasets.
Our experimental results show that the system substantially accelerates data loading by up to 21x and query execution by up to 23x and improves end-to-end performance by up to 19x within a budget of 1.0 microseconds latency per record on clients.
\end{abstract}

\section{Introduction}
Databases often centralize data collected from multiple, distributed client systems.
For example, a single log server in a data center may collect \texttt{syslog} events from all other servers~\cite{winlog}.
Or, a time-series database may collect environment sensor readings from sensors placed around a building~\cite{influxdb}.
As such deployments scale up, data loading (i.e., parsing, validating, and storing the client data) becomes an under-appreciated bottleneck on the database server due to the computation intensive tasks of parsing input formats, converting types, and validating input~\cite{adam,udp}.
This bottleneck delays the results of downstream analytics queries, and can increase the latency of any decision-making system that consumes those results. Database systems eagerly ingest and load data, as there is no mechanism for the system to determine if the data will be relevant and needed for likely future queries.

Prior research literature describes three main approaches to relieve pressure on the database: client-side parsing~\cite{SharedLoading}, raw-format query processing~\cite{sparser,nodb}, and hardware acceleration~\cite{Mison,SIMDJSON, CSVParser,Instantloading,udp}. 
In client-side parsing, we leverage excess processing capacity on the client devices to parse and serialize the data before ingestion on the central server. 
Client-side parsing reduces the data loading burden on the server and the network transfer between the server and the client. 
The main disadvantage of the client-side parsing is that it requires relatively capable and powerful client devices to implement---if the clients are too under-powered it can actually hurt the overall per-record loading latency.
An alternative is to simply avoid data loading on the server when possible, and directly process queries over the raw-format data.
While this approach avoids assumptions about the client and relieves the data loading bottleneck, 
it often results in suboptimal downstream query processing.
Structured formats, like columnar storage, may have an upfront loading cost 
but greatly improve downstream query latency especially in comparison with a row-oriented raw data format. 
Regardless, raw processing still requires expensive ingestion when a record or attribute is required for a query's predicate evaluation.

Clearly, there is a careful balancing act between client-side processing, data loading costs, and downstream query processing.
However, to an administrator, the only metric that matters is the per-record processing time: 
the time from when an event happens to when it is reflected in the query result.
The various factors combine in complex, deployment-specific ways to result in a final per-record processing time.
Today's approaches pick one point in this complex design space. 
As such, they don't give the user enough flexibility to reason about the trade-offs in a variety of hardware settings.
Client-side parsing may lead to overall worse performance when the clients are under-powered, 
and raw-format query processing may lead to worse performance if there are repeated aggregate queries over the same data.

This paper presents \sys, an optimization framework that can determine what processing to do 
on a client given a computation budget to maximally benefit downstream query processing.
\sys identifies a set of predicates that can be applied directly by the client using simple string pattern matching, 
and selects the set of predicates for a client to evaluate using a client's slackness via a specified time limit. 
Clients evaluate these predicates and include lightweight bitvectors to indicate what records satisfy what predicates. 
The server then selectively loads records that satisfy at least one predicate, 
and sets aside the other raw data to be loaded when needed (e.g. just-in-time loading). 
For records that are loaded into the internal format. \sys retains the bitvectors to use for data-skipping~\cite{data-skipping}. 
\sys is developed as part of a project on resource-efficient database systems, 
CrocodileDB~\cite{CrocDB}, to explore how to improve resource utilization in the data loading process.

The key architectural insight is a marriage between raw-format query processing and client-side parsing: 
the client devices directly manipulate the raw data without fully parsing it.
We leverage techniques similar to Sparser~\cite{sparser} and UDP~\cite{udp} to directly apply popular filters to raw data records. 
However, instead of evaluating predicates on the raw JSON data at the server, which requires complex changes to the execution engine, we use simple filtering on client-side that respects a computation budget.
These filters give us annotations that can be used for ``partial data loading'' on the server, where only the most relevant data is eagerly loaded.
The filters also facilitate data skipping when the database is queried.

The core optimization problem in \sys is to select a subset of predicates to be pushed down with respect to a computation budget on the client side.
We prove that this is a \emph{submodular} problem, that is, it has diminishing marginal returns.
We leverage algorithms from the submodular optimization literature to appropriately select what computation to do on the client with optimality guarantees.
Experimentally, we show a trade-off between the client's budget and the downstream server loading and query processing savings.
We implemented this system using popular data systems components and formats (e.g., JSON, Parquet, and Spark) 
and evaluated its performance on three real-world datasets.
Our experimental results show that the system substantially accelerates data loading by up to 23x 
and query execution by up to 21x and improves end-to-end performance by up to 19x.

The rest of the paper is organized as follows. We start in Section~\ref{sec:related} with a discussion of related work to 
position our contributions against prior work. In Section~\ref{sec:overview} we overview \sys and provide our assumptions. 
We discuss how client-side predicate evaluation works in Section~\ref{sec:client} 
and our predicate selection method in Section~\ref{sec:predicate}. 
In Section~\ref{sec:execution} we outline how the server-side uses the prefiltering information for lazy data loading and data skipping. 
We evaluate our system in Section~\ref{sec:experiments} and conclude in Section~\ref{sec:conc}.
\section{Related Work}
\label{sec:related}


We now discuss the related research projects on fast data parsing and ingestion, 
in-situ query processing with lazy data loading, and offloading work to the data clients (e.g. edge sensors). 
We note that none of these projects considers pushing predicates to the clients to 
reduce the data loading cost and accelerate query processing.

\subsection{Fast data parsing and ingestion}
Our prior study~\cite{CrocDB} and others~\cite{adam} show that data loading is a time-consuming process, 
especially for text-based data formats (e.g. JSON or CSV). 
Many research projects consider accelerating the data loading process by exploiting 
the modern hardware. 
Instant loading~\cite{Instantloading} leverages SIMD instructions to accelerate 
parsing CSV files and interleaves the index creation with parsing data 
to make data quickly available. 
Several other projects~\cite{Mison, SIMDJSON, CSVParser} exploit SIMD to quickly 
parse JSON files or general text-based data formats in a distributed setting.
UDP~\cite{udp} builds a programmable accelerator and offloads the data loading process 
to the hardware accelerator. 

\sys is different from these projects in that none of them considers 
leveraging the computing power of the clients to accelerate the data loading process. 
In addition, they do not consider partial loading to make data quickly available for queries. 

\subsection{Lazy data loading and in-situ query processing}
Many projects consider not loading the data upfront, 
but directly queries data in its raw format (e.g. CSV or JSON) 
and gradually loads data while processing queries. 
NoDB, and later RawDB,~\cite{nodb,karpathiotakis2014adaptive} executes queries over CSV files directly 
and builds light-weight indices to accelerate future queries~\cite{olma2017slalom}. 
Later work explored parallel in-situ query execution over scientific formats~\cite{blanas2014parallel}.
Invisible loading~\cite{invisible-loading} piggybacks the data loading process 
with MapReduce~\cite{dean2008mapreduce} jobs that analyze the raw data. 
Invisible loading leverages MapReduce jobs' parsing and tuple extraction operations 
to incrementally load tuples into a database system. 
Database cracking~\cite{Cracking} builds indices incrementally 
when the underlying data is queried.
A similar idea is also adopted in FishStore~\cite{Fishstore} 
that uses SIMD to support in-situ query processing on text-based data 
and incrementally constructs an index for the raw data.
Data skipping projects~\cite{data-skipping,GSOP} 
consider building coarse-grained index to skip irrelevant data. 
Specifically, for each data block (e.g. a partition of data with dozens or hundreds of MBs data size), 
they include some metadata that specifies constraints or predicates a data block meets 
(e.g. data tuples in a data block meet the constraint of \texttt{age $\leq$ 20}). 
This meta-information is used to skip data and accelerate query processing. 

The difference between \sys and these projects is that \sys considers leveraging the 
computing capacity of the data clients to build light-weight index 
to enable partial loading and accelerate query processing.

\subsection{Computation offloading}
Some prior projects in edge computing~\cite{ThinkAir, FADES} consider offloading computation work to the client-side, 
but do not apply this idea to data loading. 
Shared loading~\cite{SharedLoading} offloads the data compression phase of the data loading to the client machines. 
\sys is different from the idea that we offload the predicates to the clients 
and provide a cost-based knob to adjust the additional computation cost utilized on the client-side. 
\begin{figure*}[!t]
      \centering
       \includegraphics[height=49mm]{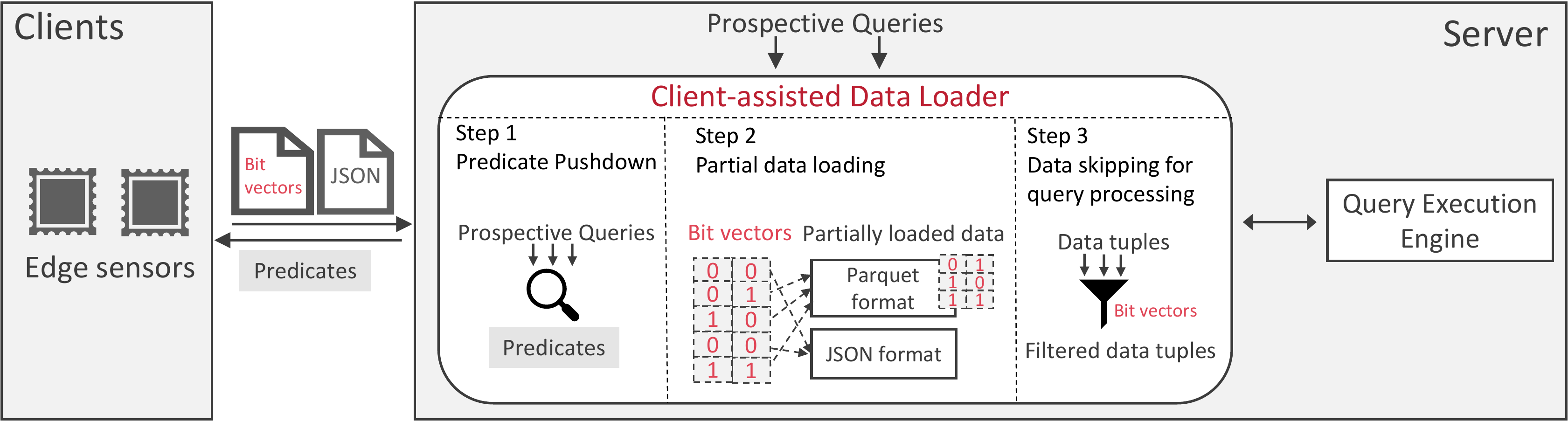}
      \caption{An overview of client-assisted data loading} 
      \label{fig:overview}
\end{figure*}

\section{Overview and Assumptions}
\label{sec:overview}

We now give an overview of \sys and show how it leverages the data clients' computing power 
to accelerate data loading and query processing. 
Specifically, \sys pushes predicates of prospective queries to the data clients (e.g. edge sensors). 
Based on a computation budget clients will evaluate simple predicates on the data before sending them to the server 
and generate bit-vectors that indicate whether a tuple is valid for a predicate. The bit-vectors are sent along with the raw data to the server. 
After, \sys utilizes the bit-vectors to selectively load the raw data format (e.g. JSON) from the clients 
into a binary data format that is more amenable to query (e.g. Parquet). 
Finally, when processing a query, the bit-vectors are used to skip irrelevant tuples that do not belong 
to the query. 

Fig.~\ref{fig:overview} shows an overview of the three steps of client-assisted data loading. 
The first step is \texttt{predicate pushdown}. 
Choosing the predicates to push down systematically considers two factors: 
how many new tuples a predicate can filter out for the prospective queries (i.e. the new tuple is marked as not valid for a predicate) 
and the increased cost of evaluating this predicate on the client-side. 
Therefore, this step takes the following information as the input to decide the predicates to be pushed down: 
1) the frequencies of queries that are expected to be executed;
2) the selectivity of each predicate in the prospective queries; 
3) the cost of evaluating a predicate on the client-side; 
and 4) a \textit{computation budget} that we allow on the client-side to evaluate the predicates we choose to push down. 
Here the computation budget is specified by the database administrator
and is defined as the average amount of computation cost of evaluating predicates for each new tuple. 
We estimate the frequencies of prospective queries and selectivities of predicates based on historical statistics. 
We develop a cost model to estimate the cost of evaluating a predicate, which is shown in Sec.~\ref{sec:costmodel}. 

Given this information, we use a greedy algorithm to decide 
the predicates to be pushed down to the client-side with respect to the computation budget. 
This algorithm is optimized to select the predicates that filter out the most tuples 
for each unit of increased computation cost on the client-side, which is discussed in Sec.~\ref{sec:predicate}. 
In this paper, we assume that the data clients  generate JSON objects 
and use string operations (e.g. check whether a JSON object contains a substring) 
to evaluate the predicates. 
Sec.~\ref{sec:problemsetup} discusses the predicates we support. 
We further assume that data clients send JSON objects in chunks (e.g. 1k objects for each chunk). 
Each \textit{JSON chunk} is associated with a set of bit-vectors, where each bit-vector corresponds to a predicate. 
As shown in Step 2 of Fig.~\ref{fig:overview}, a bit 0 means the tuple is invalid to a predicate; 
otherwise, the bit is 1.

When \sys receives the JSON chunks from the clients, we selectively load the parts of the JSON objects (i.e. data tuples) into Parquet files, 
which is shown as Step 2 in Fig.~\ref{fig:overview}. 
Specifically, we choose to load a JSON object if it is at least valid to one predicate, 
that is, the JSON object's bit is marked as valid for at least one bit-vector. 
The rationale here is that we load the data that is likely to be accessed by prospective queries. 
Therefore, a JSON chunk is split into two parts: one is loaded into the Parquet format that is available for querying 
and the other is left in a raw JSON format, which requires later parsing and conversion to analyze the unprocessed records. 
The parquet file is also associated with a set of new bit-vectors that are derived from the bit-vectors of the original JSON chunk 
and represent whether a tuple is valid for each predicate. 
During query processing, we leverage these bit-vectors to accelerate query processing (i.e. Step 3 in Fig.~\ref{fig:overview}). 
We discuss Step 2 and 3 in Sec.~\ref{sec:execution}. 

\section{Client-side Predicate Evaluation}
\label{sec:client}
A core contribution of our framework is to \emph{evaluate query predicates on client-devices without full parsing}.
We argue that this design decision achieves the best of both worlds: it reduces data loading costs on the server, while not shifting the parsing burden to the clients.

\subsection{Raw-data format}
Data acquired from client systems are
often generated in a string-based raw-data format.
Such formats, like delimited files and JSON (JavaScript Object Notation), are highly portable across architectures and programming languages.
Being strings such data are also easy to profile and debug.
These formats are also highly general as they can represent many different data types and both flat and nested structures.
However, this generality means that the downstream parser has to expend additional computation for parsing and validation to support features the user may not use (e.g., like parsing escape characters).

Thus, a natural question studied in a number of recent works is whether it is possible to avoid full parsing for specific types of queries~\cite{sparser,nodb,invisible-loading}.
Any given predicate may not require a full, structured representation of the raw data.
Our twist on this problem is to consider a client-server extension to this basic idea of raw-format query processing to facilitate both data-skipping on the server and partial data loading to avoid loading irrelevant data.

For simplicity, we assume that the client-side generates data tuples in the JSON  format. 
We note our solution can also be applied to other text-based data formats, like CSV.
JSON is a common data exchanging format and a JSON file is stored as human-readable text.
Each JSON file is composed of a set of JSON objects, where each object is defined as a set of key-value pairs: 
\texttt{Object=\{String: Value, String: Value, $\cdots$, String: Value\}}. 
A JSON value could be a string, number, null, a boolean value (i.e. true/false), 
a recursive JSON object, or an array. 
An array is defined as \texttt{Array=[Value, $\cdots$, Value]}.
One JSON example could be \texttt{\{``name'':``Bob'', ``age'':22\}},
where we have two keys \texttt{``name''} and \texttt{``age''}, 
with two values \texttt{``Bod''} and \texttt{22} respectively.

\subsection{String-based predicate evaluation}
Since JSON objects are represented as strings, a limited number of SQL predicates can be evaluated as string search operations.
Our client-side framework converts supported SQL predicates into string-based pattern expressions that can quickly identify satisfying JSON objects.
We currently support the following types of predicates 
and the corresponding examples are shown in Table~\ref{tbl:supported-predicates}
\begin{itemize}
    \item \textbf{Exact or Substring Match} 
    The first two examples in Table~\ref{tbl:supported-predicates} show the exact match 
    and substring match respectively. 
    For the exact match, the pattern string is the operand string (e.g. \texttt{``Bob''}) that compares against the value of a key.
    For the substring match, the pattern string is the substring we need to find (e.g. \texttt{``delicious''}). 
    \item \textbf{Key-presence match}
    For key-presence match, the pattern string is the key string (e.g. the \texttt{``email''} of the third row in Table~\ref{tbl:supported-predicates}). 
    \item \textbf{Key-value match} 
    For key-value match, we have two pattern strings: the key string and the value string. 
    The client first searches for the key string 
    and if the key string exists, the client will continue from the current string position to search for the next key-value delimiter 
    (i.e. a comma \texttt{``,''}).
    Between the key string and delimiter, the client will check whether the value string exists. 
\end{itemize}

Client-side filtering is a sort of converse to data-prefetching---as it is a bet that the server can make to hide latency from the end-user.
And, similar to pre-fetching, we engineer client-side filtering to allow for false-positive cases, that is,
a JSON object that is actually not valid for a predicate can be marked as valid.

Consider the example of exact match in Table~\ref{tbl:supported-predicates}. 
The pattern string \texttt{``Bob''} can exist in the key-value pair that does not include \texttt{``name''} as the key. 
Therefore, when a query scans the filtered tuples based on the bit-vectors, 
it needs to evaluate all predicates in this query to verify that a tuple is actually valid to this query. 
However, false-negative cases will never happen in our predicate evaluation, 
that is, if we cannot find the pattern strings in a JSON object, this JSON object is not valid to the corresponding predicate. 

False-negative cases are not allowed in the predicate evaluation because they discard the JSON objects 
that should have been incorporated into query results. 
Therefore, range-based predicates or inequality predicates are not supported. 
In addition, if there are different data representations for the same number (\texttt{2.4} vs. \texttt{24e-1}),
we do not support the number equality 
because it will result in false negative cases. 
These limitations also exist in other systems that evaluate predicates on Raw JSON objects (e.g. Sparser\cite{sparser}).

We implement the client-side framework in C++. 
We use the \textit{string::find} method of C++ STL for substring matching.

\begin{table}[!t]
    \caption{Examples for supported predicates and pattern strings}
    \label{tbl:supported-predicates}
    \begin{tabular}{@{}clc@{}}
    \toprule
    \small
    \textbf{Supported Predicates} & \textbf{Example}          & \textbf{Pattern String}                              \\ \midrule
    Exact String Match            & name = ``Bob''              & ``Bob''                                                \\
    Substring Match               & text LIKE ``\%delicious\%'' & ``delicious''                                          \\
    Key-Presence Match            &  email != NULL            & ``email''                                              \\
    Key-Value Match               & age = 10                  & ``age'' ``10''\\ \bottomrule
    \end{tabular}
    \end{table}

\section{Predicate Selection Optimization}
\label{sec:predicate}
Each predicate evaluated on the client incurs a cost.
Since client-devices are often under-powered compared to the server, these costs can be significant.
In this section, we describe an optimization algorithm for selecting which predicates to push down to clients.
We can formulate this problem as a submodular maximization problem. 
Such problems are, not surprisingly, the class of problems with diminishing marginal returns, and are relevant because each additional predicate that is pushed down has diminishing returns due to overlaps.
Such ideas have been leveraged in data management optimization problems in a number of prior works~\cite{choenni1993selection,li2018submodularity}.

We first set up the optimization problem in Sec.~\ref{sec:problemsetup} 
and then show the key property \texttt{submodularity} for this problem in Sec.~\ref{sec:submodular}. 
After, based on the submodularity, we use an approximation algorithm in Sec.~\ref{sec:greedy} 
to solve this optimization problem. 
Finally, we discuss estimating the cost of evaluating predicates on the client-side in Sec.~\ref{sec:costmodel}.

\subsection{Problem setup and cost model}
\label{sec:problemsetup}
Consider a workload of $m$ queries $Q = \{q_1, q_2, \cdots , q_m \}$, where each query has a predicate that is a \emph{conjunction of disjunctive clauses}.
For example, $q_i$ has the two conjunctive clauses: \texttt{$q_i$: name in (``Bob'', ``John'') and age = 20}. 
Here, \texttt{name in (``Bob'', ``John'')} is a disjunction (i.e. \texttt{name = ``Bob'' or name = ``John''}). 
When we choose the predicates to push down, we consider each conjunctive clause as an atomic unit (each clause is hereafter referred to as a ``predicate'').
For example, simply pushing down \texttt{name = ``Bob''} cannot discard tuples for the predicate \texttt{name in (``Bob'', ``John'')}.

Thus, for every query in the workload we have a set of candidate predicates to pushdown: for $q_i$ the set of clauses (predicates) $P_i = \{p_1, p_2, \cdots , p_{n_i}\}$ that can be evaluated on the client. 
If a conjunctive predicate includes a disjunctive clause that we cannot support on the client, we do not consider it as a candidate to push down.

We further define some valuable notation that will help us formalize the optimization problem.
Let the selectivity of a predicate $p$ be denoted as $sel(p)$ and 
the cost of evaluating $p$ for a JSON object as $cost(p)$.
Let $freq(q)$ denote an estimate of the relative frequency that $q$ is evaluated.
We further assume that the computation budget on the client-side is fixed cost $B$, where $B$ represents the average units of computation cost of evaluating predicates for each new tuple. 

Let $S$ be the set of predicates that we chose to evaluate on the client. 
For each query $q_i$, let $S_i = (P_i\cap S)$ be the set of the query's conjunctive clauses that have been pushed down. 
We can evaluate the probability of filtering a JSON object given the selected predicate set $S$ 
is using a statistical independence assumption and the selectivity of each predicate:
\[
f(q_i, S)  = 1-\prod_{p_j\in (S_i)} sel(p_j)
\]
The optimization goal here is to maximize the \emph{expected benefit} of predicate pushdown for all queries in $Q$:
\[
f(S) =  \sum_{q \in Q} f(q, S) \cdot freq(q).
\]
We want to optimize this quantity while ensuring that $\sum_{p_i \in S} cost(p_i) \le B$.
In our experiments and the rest of the text, we present results with a uniform query frequency (though not necessarily a uniform predicate frequency).
Formally, the optimization problem is defined as:
\begin{equation*}
\begin{aligned}
    & \text{maximize} & & f(S) \\
    & \text{subject to} & &  \sum_{p_i \in S} cost(p_i) \le B
\end{aligned}
\end{equation*}

\renewcommand{\algorithmicrequire}{\textbf{Input:}}
\renewcommand{\algorithmicensure}{\textbf{Output:}}
\removelatexerror
\begin{algorithm}[!t]
    \caption{Naive greedy algorithm}
    \begin{algorithmic}[1]
        \REQUIRE The set of conjunctive predicates $P$ for all queries
        \ENSURE Selected predicate set $S$  
        \STATE Let $S \leftarrow \emptyset$.
        \WHILE {$\exists p \in P\setminus S: cost(p) + \sum_{p_i \in S} cost(p_i) \le B$}
        \STATE let $S \leftarrow S \cup \left\{ \mathop{\arg\max}\limits_{p \in P\setminus S} f(S\cup\{p\}) \right\}$.
        \ENDWHILE
        \STATE Return $S$.
    \end{algorithmic}
    \label{alg:naive}
\end{algorithm}

\subsection{Submodularity}
\label{sec:submodular}
We now show that $f(S)$ is a submodular set function and based on the submodularity 
we use an approximation algorithm to solve the optimization problem (in Sec.~\ref{sec:greedy}). 
By definition, $f(\cdot)$ is a submodular set function if 
for any two predicate sets $S$ and $T$, $f(S) + f(T) \geq f(S\cap T) + f(S\cup T)$.
Recall that $f(S) = \sum_{i=1}^{m} [1-\prod_{p_j\in S_i}sel(p_j)]$. 
Therefore, if $f(q_i, S) = 1-\prod_{p_j\in S_i}sel(p_j)$ is a submodular function for any query $q_i$, 
then $f(S)$ is also a submodular function. 
For simplicity, we let $g(S) = f(q_i, S)$

Given a query $q_i$, we now show that $g(S) + g(T) - g(S\cap T) - g(S\cup T) \geq 0$. 
Assuming that $g(S) = 1 - h(S)$, where $h(S) = \prod_{p_j\in (P_i\cap S)}sel(p_j)$, 
we have
\begin{equation*}
    \begin{aligned}
    g(S) + & g(T) - g(S\cap T) - g(S\cup T) \\
    & = h(S\cap T) + h(S\cup T) - h(S) - h(T) \\
    & = h(S\cap T) \times [1 + \frac{h(S \cup T)}{h(S\cap T)} - \frac{h(S)}{h(S\cap T)} - \frac{h(T)}{h(S\cap T)}] \\
    & = h(S\cap T) \times [1 + h(S^{\prime})\times h(T^{\prime}) - h(S^{\prime}) - h(T^{\prime})] \\
    & = h(S\cap T) \times [1 - h(S^{\prime})] \times [1 - h(T^{\prime})]
    \end{aligned}
\end{equation*}
where $S^{\prime} = S \setminus (S \cap T)$ and $T^{\prime} = T \setminus (S \cap T)$. 
Since $h(S)$ is the product of the selectivities of the predicates in $S$ for $q_i$, 
$\frac{h(S)}{h(S\cap T)}$ essentially removes the predicates in $S\cap T$ for $S$. 
Therefore $\frac{h(S)}{h(S\cap T)} = h(S^{\prime})$. 
Similarly, $\frac{h(T)}{h(S\cap T)} = h(T^{\prime})$ and $ \frac{h(S \cup T)}{h(S\cap T)} =  h(S^{\prime})\times h(T^{\prime})$.
Since the value of $h(\cdot)$ is in the range $[0, 1]$, the above equation is no smaller than 0, 
which means that $g(S)$ is a submodular function and so is $f(S)$.

\begin{algorithm}[!t]
    \caption{Greedy algorithm based on benefit-cost ratio}
    \begin{algorithmic}[1]
        \REQUIRE The set of conjunctive predicates $P$ for all queries
        \ENSURE Selected predicate set $S$  
        \STATE Let $S \leftarrow \emptyset$.
        \WHILE {$\exists p \in P\setminus S: cost(p) + \sum_{p_i \in S} cost(p_i) \le B$}
        \STATE $S \leftarrow S \cup \left\{ \mathop{\arg\max}\limits_{p \in P\setminus S} \frac{f(S\cup\{p\})-f(S)}{cost(p)} \right\}$.
        \ENDWHILE
        \STATE Return $S$.
    \end{algorithmic}
    \label{alg:ratio}
\end{algorithm}

\subsection{Approximation algorithm}
\label{sec:greedy}
As the previous work~\cite{submodularSurvey} shows, this is a submodular maximization problem 
with a knapsack constraint (i.e. $\sum_{p_i \in S} cost(p_i) \le B$). 
One naive greedy algorithm is to repeatedly add one predicate to a selected predicate set 
and at each step choose the predicate that mostly increases the optimization goal $f(S)$, 
which is shown in Algorithm~\ref{alg:naive}. 
It first checks whether there exists one unselected predicate $p$ 
which does not break the budget constraint if $p$ is added to the current selected predicate set $S$.
Then, the algorithm selects the predicate that yields the highest benefit, 
that is, $\mathop{\arg\max}\limits_{p \in P\setminus S} f(S\cup\{p\})$. 
However, this algorithm can perform arbitrarily badly 
since it does not consider the cost of evaluating a predicate when it chooses the predicates~\cite{submodularSurvey}. 
A variant algorithm is to choose the predicate that yields the highest benefit-cost ratio, 
that is, for a predicate $p$ it computes the benefit-cost ratio as $\frac{f(S\cup\{p\})-f(S)}{cost(p)}$, 
which is shown in Algorithm~\ref{alg:ratio}. 
The prior work~\cite{submodularSurvey} shows that this solution can also perform arbitrary badly with respect to the optimal solution. 

Fortunately, the better solution of the two algorithms has a bounded error with respect to the optimal solution. 
Specifically, we run the two algorithms separately and choose the one with the higher $f(S)$. 
One prior study~\cite{boundAprrox} proves that this solution is no smaller than 
$\frac{1}{2} (1 - \frac{1}{e}) OPT \approx 0.316\times OPT$ where $e$ is the mathematical constant 
and $OPT$ is the $f(S)$ of the optimal solution. 

\subsection{Cost model for predicate evaluation}
\label{sec:costmodel}
We now discuss how to estimate the cost of evaluating a predicate on the client-side.
For a disjunction of predicates (e.g. \texttt{name = ``Bob'' or age = 10}), 
its cost is the summation of the cost of evaluating each simple predicate. 
Therefore, our following discussion is focused on estimating the cost of evaluating a simple predicate.
We find that the substring match operation is the basic operation that implements 
the evaluation of all predicates that we support. 
We now discuss the cost estimation of the substring match cost.

We perform some experiments to understand the key factors that influence the cost of a substring search. 
We find that the time cost is proportional to pattern string length and the length of a JSON object. 
We also observe that the cost for the case that a pattern string is found in a JSON object 
is different from the otherwise case. 
Therefore, we model the two cases differently.

We collect information such as the average JSON object length of a dataset from historical statistics.
The expected cost of evaluating a predicate $p$ on one JSON object is modeled as:
\begin{equation*}
\begin{aligned}
T = & sel(p) \times [k_1 \times len(p) + k_2 \times len(t)] \\
    & + [1 - sel(p)] \times [k_3 \times len(p) + k_4 \times len(t)] + c
\end{aligned}
\end{equation*}
where $sel(p)$ denotes the selectivity of predicate $p$, 
$len(p)$ represents the pattern string length, 
$len(t)$ indicates the average JSON object length. 
The first term in the equation models the cost when a pattern string is found in a JSON object 
and the second term corresponds to the cost when a JSON object does not have a match of the pattern string. 
The third term $c$ is the startup cost for each substring search.
The constants $k_1$, $k_2$, $k_3$, $k_4$ and $c$ are dependent on the hardware configuration 
and are estimated from historical statistics.
We evaluate this cost model in Sec.~\ref{sec:sensitivity_costmodel}. 
\section{Partial Data Loading and Data Skipping}
\label{sec:execution}

After we select a subset of predicates, we associate each predicate with an \texttt{id} 
and generate its pattern strings. 
We use a \texttt{predicate hashmap} to store this information. 
It uses the predicate as the key and the predicate id and its pattern strings as the value. 
Fig.~\ref{fig:workflow} shows an example of the predicate mapping. 
The predicate \texttt{name = ``Bob''} has an id \texttt{1} and the pattern string \texttt{``Bob''}.
\sys pushes down the pattern strings along with the predicate ids to the clients. 
After, the clients will use substring (e.g., simple pattern matching) operations to evaluate predicates on each JSON object 
and generate bit-vectors that can be used by the server to partially load data 
and skip irrelevant tuples for query processing. 
In the following discussion, we use the example in Fig.~\ref{fig:workflow} 
to explain partial data loading and data skipping in Sec.~\ref{sec:partial} 
and Sec.~\ref{sec:skipping} respectively. 
After, we discuss how the query workload impacts the performance of \sys.
We use the example in Fig.~\ref{fig:workflow} to explain the workflow of \sys.

\subsection{Partial data loading}
\label{sec:partial}

\begin{figure}[!t]
      \centering
       \includegraphics[height=49mm]{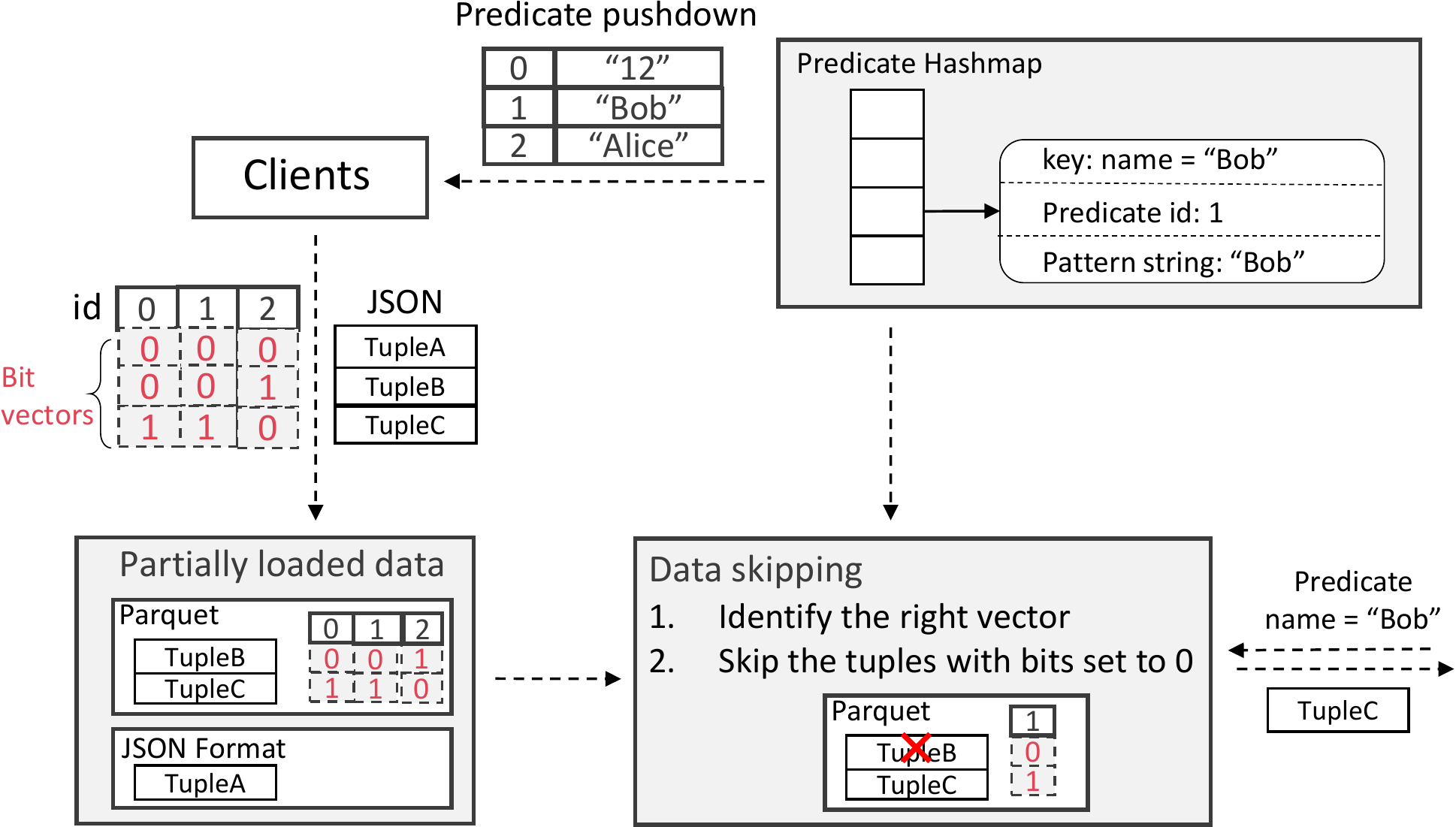}
      \caption{An example of the workflow of \sys} 
      \label{fig:workflow}
\end{figure}

The data clients send JSON chunks to the server. 
Each JSON chunk includes the JSON objects (i.e. data tuples) and a set of bit-vectors, 
where each bit-vector corresponds to a predicate.
Fig.~\ref{fig:workflow} shows the bit-vectors sent from the clients 
and these bit-vectors can be indexed by the predicate ids.
For each JSON chunk, the server loads parts of the JSON objects 
into Parquet, a data format that is faster for queries to access, 
and leaves the rest of the JSON objects unloaded.
The intuition of partial data loading is that 
data loading is a time-consuming process as shown in prior work~\cite{adam} 
and we choose to not load the data that is unlikely to be accessed by prospectively queries.
Specifically, for each JSON object, if its associated bits of all predicates are 0 (i.e. not valid for all predicates pushed down), 
this JSON object is unlikely to be accessed and is not loaded into Parquet format. 
Otherwise, the tuple is loaded into a Parquet file.
For example, Fig.~\ref{fig:workflow} shows that \texttt{TupleB} and \texttt{TupleC} are converted into Parquet format, 
but \texttt{TupleA} is left as JSON format.
When we load a JSON object into a Parquet file, we store the bit-vector information of this object 
into the metadata of each data block of the Parquet file.
The bit-vectors are then used for data skipping when processing new queries.

\subsection{Data skipping}
\label{sec:skipping}

When a query is submitted to the database and scans the data tuples from \sys, 
\sys first extracts the conjunctive predicates from the query and compares them 
to the predicates we have pushed down. 
If we find the query includes predicates that we have pushed down, 
we only need to scan the Parquet file 
because the unprocessed JSON files do not include any tuples that are valid to the predicates we have pushed down. 
When we scan data blocks of the Parquet file, 
we extract the bit-vectors that belong to the predicates included in the query,
take the intersected bit-vector of them (i.e. using AND due to the conjunctive predicates), 
and use the bit-vector to skip data tuples. 
For each tuple from a data block, if the corresponding bit of the intersected bit-vector is 0 
(i.e. not valid for at least one conjunctive predicate),
we discard it. Otherwise, we output it to the query process engine. 
For example, Fig.~\ref{fig:workflow} shows a query with a predicate \texttt{name = ``Bob''}. 
We find its predicate id in the predicate hashmap (i.e. id = 1) 
and filter out the tuples whose corresponding bits are set to 0 (i.e. discarding \texttt{TupleA} and \texttt{TupleB}). 
When the query does not include a predicate that we have pushed down, \sys scans both Parquet and 
JSON files to return all data tuples.

%
%
%

\subsection{Performance impact of a query workload}
\label{sec:perf-workload}

We find the performance of \sys depends on the following factors of a query workload. 
We break down the performance impact of these factors in the micro benchmark experiments in Sec.~\ref{sec:micro}. 

\stitle{Predicate overlap} 
Predicate overlap is one significant factor that impacts system performance.
If a large number of predicates are shared by many queries, 
pushing them down will be beneficial for many queries. 
On the other hand, however, if there is no predicate overlap across queries 
(i.e. all queries have a distinct set of predicates), 
the bit-vector generated for one predicate is only useful for skipping tuples for one query. 
In addition, partial loading likely loads more data in this case. 
To achieve the same query performance (e.g. query latency), 
we need to set a higher budget for the query workload that has smaller predicate overlap.

\stitle{Predicate selectivity} 
The predicate selectivity also impacts the number of tuples we need to load and the effectiveness of data skipping. 
A highly selective predicate (i.e. with low selectivity) filters more tuples 
and thus is more effective for skipping irrelevant data. 
In addition, the union of selectivities of all predicates we have pushed down 
decides the number of tuples we need to load. 
If we have a predicate has a lower selectivity, 
we are likely to load less data.

\stitle{Predicate skewness}
Predicate skewness represents the skewness of the probability of one predicate appearing in one query. 
If there are a small number of predicates that exist in many queries, 
pushing these predicates down is enough to cover all queries (i.e. a query includes at least one of the predicates that are pushed down). 
Therefore, a skewer distribution of the predicates across prospective queries 
is more beneficial for partial data loading and data skipping, 
and consumes a smaller computation cost on the client-side.
However, if predicates are uniformly distributed, we need to push down a larger number of predicates 
to achieve a similar performance compared to the skew distribution. 

\section{Experiment Results}
\label{sec:experiments}
We evaluate CIAO on three axes:

\begin{itemize}
    \item How effective is \sys in reducing the data loading time and query execution time with varied computation budgets for different workloads (in Sec.~\ref{sec:end-to-end})?
    \item How does the predicate overlap, selectivity, and skewness impact the performance of \sys (in Sec.~\ref{sec:micro})?
    \item How robust is our cost model across different hardware configurations (in Sec.~\ref{sec:sensitivity_costmodel})?
\end{itemize}

\subsection{Prototype implementation}
We implement the predicate selection algorithm in Python 3, 
and the components of evaluating predicates on the client-side and partial data loading in C++. 
We use the \texttt{string::find} method of C++ STL for substring matching 
and choose rapidJSON\cite{rapidjson} as our JSON parser when we partially convert JSON objects to the parquet file.
We build parquet files with the low-level interfaces provided by the Apache Arrow C++ project\cite{arrow}.
We integrate our data skipping mechanism with the query execution engine of Apache Spark 2.4 
by checking corresponding bit vectors to decide whether to discard a tuple. 

All experiments are conducted in a single-node Linux machine with a 2-core Intel Core i7-5557U CPU @ 3.10GHz and 16 GB RAM.
To simulate a server-client deployment, we implement a single-client and server on the same machine. All communication is simulated through file I/O, and all of the experiment processes are single-threaded.

\subsection{Datasets}
We evaluate \sys using three JSON datasets.

\stitle{Yelp Open Dataset}: 
The Yelp open dataset\cite{yelp} is released by Yelp Inc., a company that publishes crowd-sourced reviews about businesses. 
The dataset is in JSON format and each line contains one JSON-object. 
The entire dataset contains 6 files including 1.3 million tips by 1.9 million users and 1.4 million business attributes like hours, parking, etc. 
In our experiment, we use the 5 GB \textit{review.json} file which includes 6,1685,900 JSON objects. 
Each object contains the full review text data as well as the userId, 
the businessId, the date of review, and values of 4 other metrics a reviewer can use to evaluate a business.

\stitle{Windows System Log}:
The Windows System Log dataset\cite{winlog} is collected on a Windows 7 machine. 
The text file contains 114 millions of rows. Each row contains the date and time of the log, the level of the log, 
the Windows service that generates the log as well as the log message. 
The uncompressed file is 27G and spans 226 days.

\stitle{Yahoo Cloud Serving Benchmark}:
The Yahoo Cloud Serving Benchmark (YCSB)~\cite{ycsb} is a framework that can be used to evaluate different key-value and cloud services. 
It comprises two main components: the core workloads and a workload generator. 
In our experiment, we used an open-source JSON data generator \textit{fakeit}\cite{fakeit} to generate a JSON dataset
of customers which include 25 attributes, such as name, children, address, phone, email, visited places, etc.
We generate a 20 GB JSON file with 14.4 million objects for the experiments.

\begin{table}[t]
\caption{Predicate templates and the number of candidate values for each predicate template}
\begin{tabular}{@{}ccc@{}}
\toprule
\textbf{Dataset}                    & \textbf{Predicate Template}                      & \textbf{\#Candidates} \\ \midrule
\multirow{8}{*}{Yelp review}        & useful = \textless{}int\textgreater{}            & 100                        \\
                                    & cool = \textless{}int\textgreater{}              & 100                        \\
                                    & funny = \textless{}int\textgreater{}             & 100                        \\
                                    & stars =  \textless{}int\textgreater{}                             & 5                          \\
                                    & user\_id = \textless{}string\textgreater{}       & 5                          \\
                                    & text LIKE \textless{}string\textgreater{}        & 5                          \\
                                    & date LIKE "\%20{[}0-1{]}{[}0-9{]}\%" (year)      & 14                         \\
                                    & date LIKE "\%-{[}0-1{]}{[}0-9{]}-\%" (month)     & 12                         \\ \midrule
\multirow{6}{*}{Windows log} & info LIKE \textless{}string\textgreater{}        & 200                        \\
                                    & time LIKE "\%-{[}0-1{]}{[}0-9{]}-\%" (month)     & 12                         \\
                                    & time LIKE "\%-{[}0-3{]}{[}0-9{]} \%" (day)       & 31                         \\
                                    & time LIKE "\%{[}0-2{]}{[}0-9{]}:\%" (hour)       & 24                         \\
                                    & time LIKE "\%:{[}0-5{]}{[}0-9{]}:\%" (minute)    & 60                         \\
                                    & time LIKE "\%:{[}0-5{]}{[}0-9{]},\%" (second)    & 60                         \\ \midrule
\multirow{9}{*}{YCSB}               & isActive = \textless{}boolean\textgreater{}      & 2                          \\
                                    & linear\_score = \textless{}int\textgreater{}     & 100                        \\
                                    & weighted\_score = \textless{}int\textgreater{}   & 100                        \\
                                    & phone\_country = \textless{}string\textgreater{} & 3                          \\
                                    & age\_group = \textless{}string\textgreater{}     & 4                          \\
                                    & age\_by\_group = \textless{}int\textgreater{}    & 100                        \\
                                    & url\_domain LIKE \textless{}string\textgreater{} & 12                         \\
                                    & url\_site LIKE \textless{}string\textgreater{}   & 14                         \\
                                    & email LIKE \textless{}string\textgreater{}       & 2                          \\ \bottomrule
\end{tabular}
\label{tab:predicate}
\end{table}

\begin{table}[!t]
\caption{Workloads for End-to-End Experiments}
\label{tab:workload}
\centering
\begin{tabular}{@{}cccc@{}}
\toprule
\textbf{Workload} & \textbf{\#Predicates} & \textbf{\begin{tabular}[c]{@{}c@{}}Min/Max \\ \#Predicates\end{tabular}} & \textbf{\begin{tabular}[c]{@{}c@{}}Predicate \\ Distribution\end{tabular}} \\ \midrule
A                 & 732                   & 1/8                                                                                  & Zipfian(1.5)                                                               \\
B                 & 617                   & 1/7                                                                                  & Zipfian(2)                                                                 \\
C                 & 607                   & 1/10                                                                                 & Uniform                                                                    \\ \bottomrule
\end{tabular}
\end{table}

\subsection{Synthetic query workloads}
We generate the query workloads for different datasets using a single query template: 
\texttt{SELECT COUNT(*) FROM <dataset> WHERE <conjunctive predicates>}. 
We use this query template because it best evaluates the cost of scanning tuples from the base tables and shows the performance impact of partial data loading and data skipping. 
To generate queries for each dataset, 
we build a predicate pool and randomly draw the predicates from the pool to build each query's \texttt{conjunctive predicates}. 
We generate the predicate pools using predicate templates and change the candidate values for each predicate template. 
Table~\ref{tab:predicate} shows all predicate templates and the number of candidate values for each template.
For example, the template \texttt{stars = \textless{}int\textgreater{}} for the Yelp dataset in Table~\ref{tab:predicate} 
has 5 candidate values and we have five candidate predicates for this template.
We estimate the selectivity for each predicate by evaluating them on sampled datasets.

To generate the conjunctive predicates of each query, 
we assign each predicate a probability that indicates whether this predicate is selected from the pool or not. 
We make sure that each query includes the same expected number of predicates 
and vary the distribution of how likely a predicate is selected. 
For example, if the predicate pool size is 100, the expected number of predicates is 3, and we use a uniform distribution, 
each predicate will be selected with the probability $\frac{3}{100} = 0.03$. 
It is possible to use a skewed distribution, like Zipfian, 
to simulate different levels of predicate skewness 
(i.e. if the distribution is more skew, a small number of predicates have higher probabilities of appearing in a query).
In the following experiments, we set the expected number of predicates for each query to 3 unless otherwise specified and vary the distributions of selecting the predicates to simulate different workloads.


\begin{figure*}[!t]
      \centering
       \includegraphics[height=40mm]{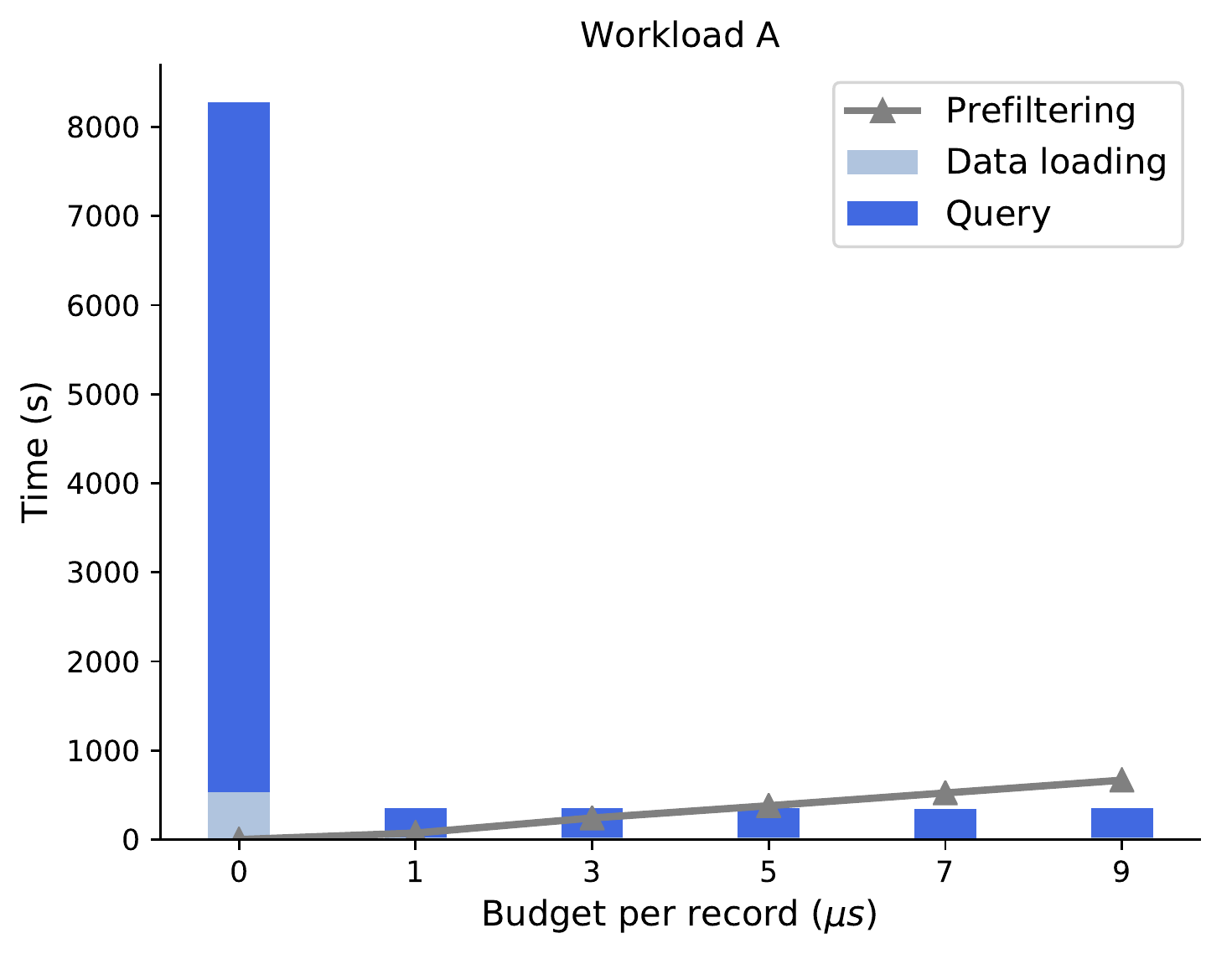}
       \includegraphics[height=40mm]{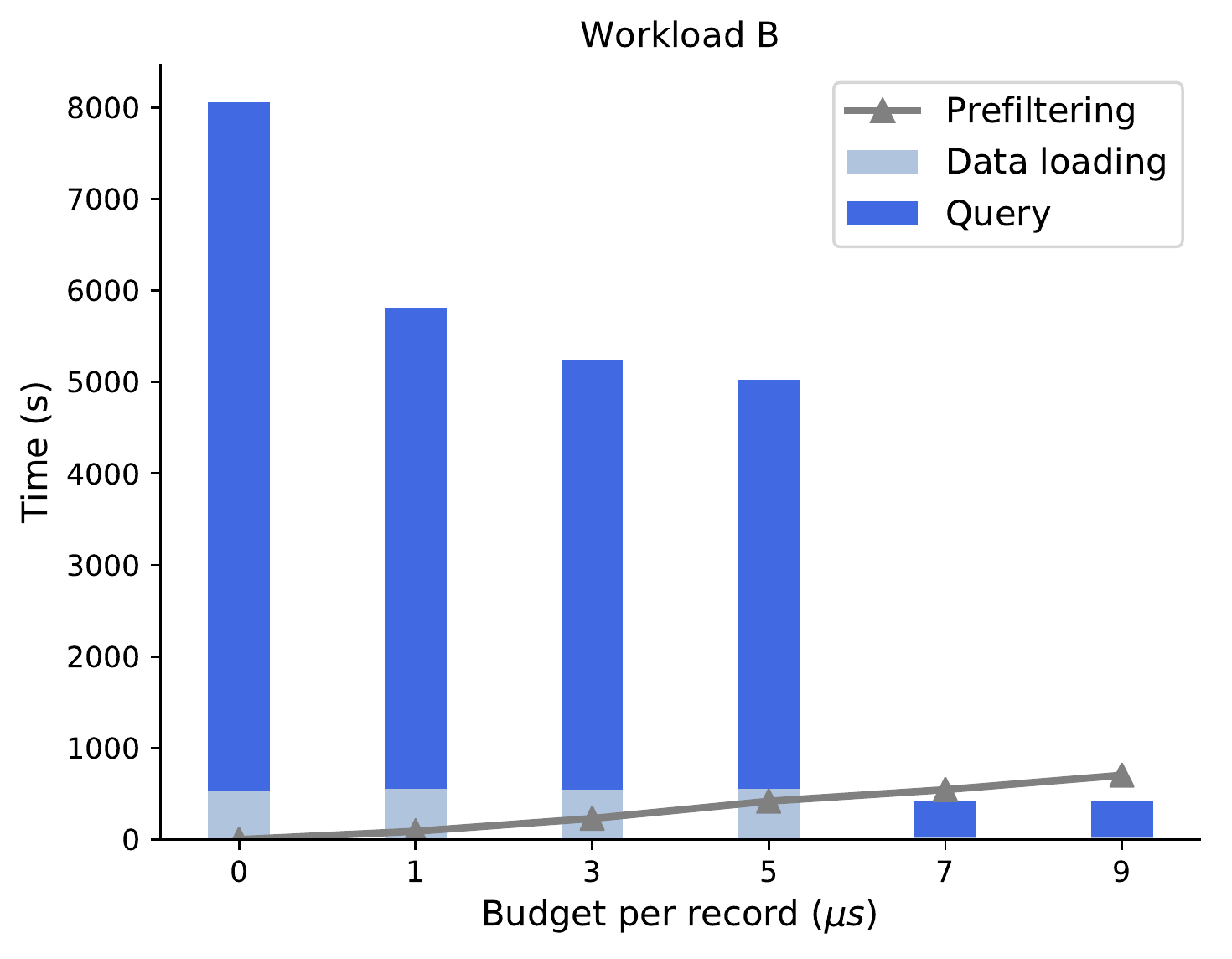}
       \includegraphics[height=40mm]{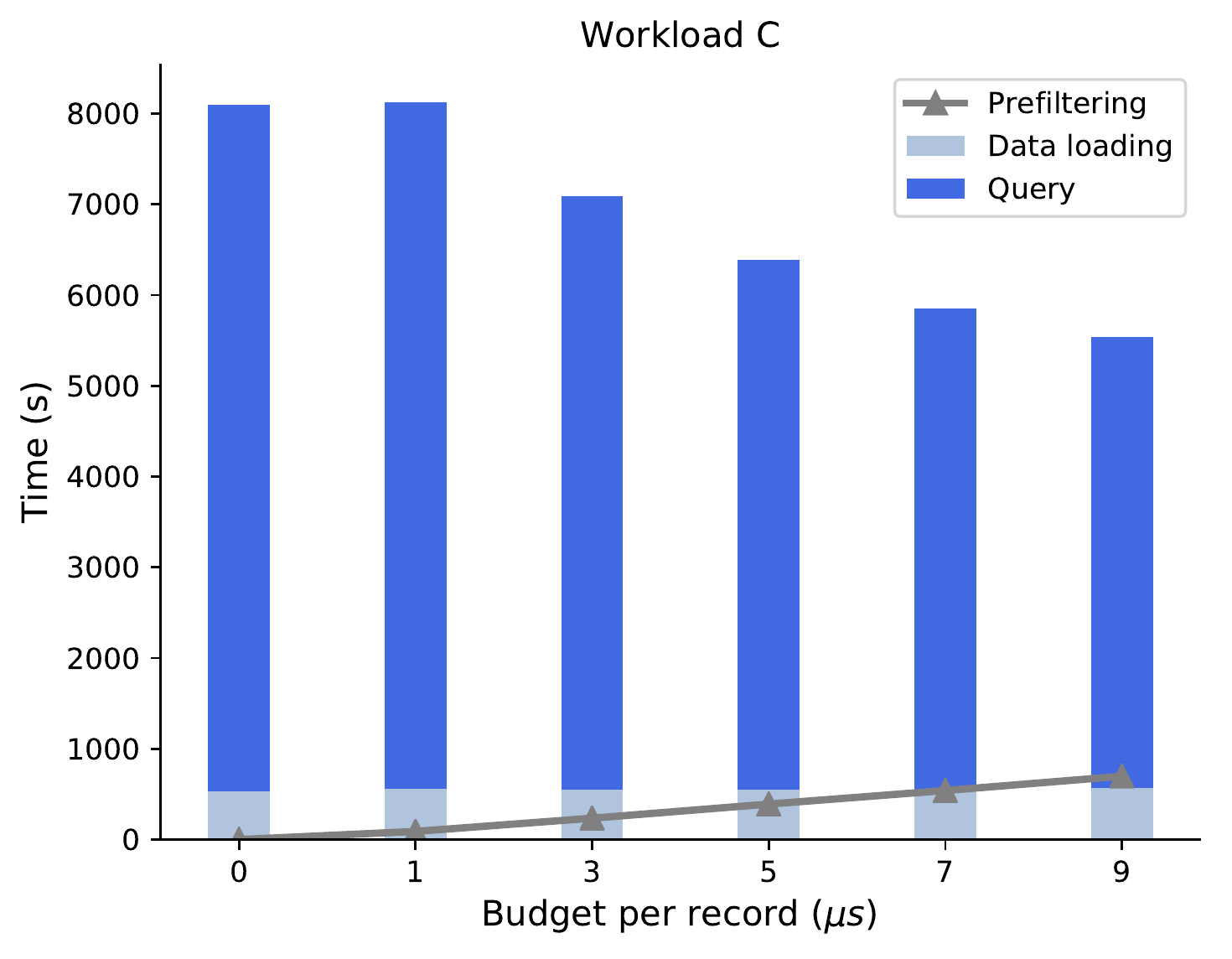}
      \caption{End-to-End Experiments of the 3 workloads on the Windows System Log Dataset}
      \label{fig:end2end-win}
\end{figure*}

\begin{figure*}[!t]
      \centering
       \includegraphics[height=40mm]{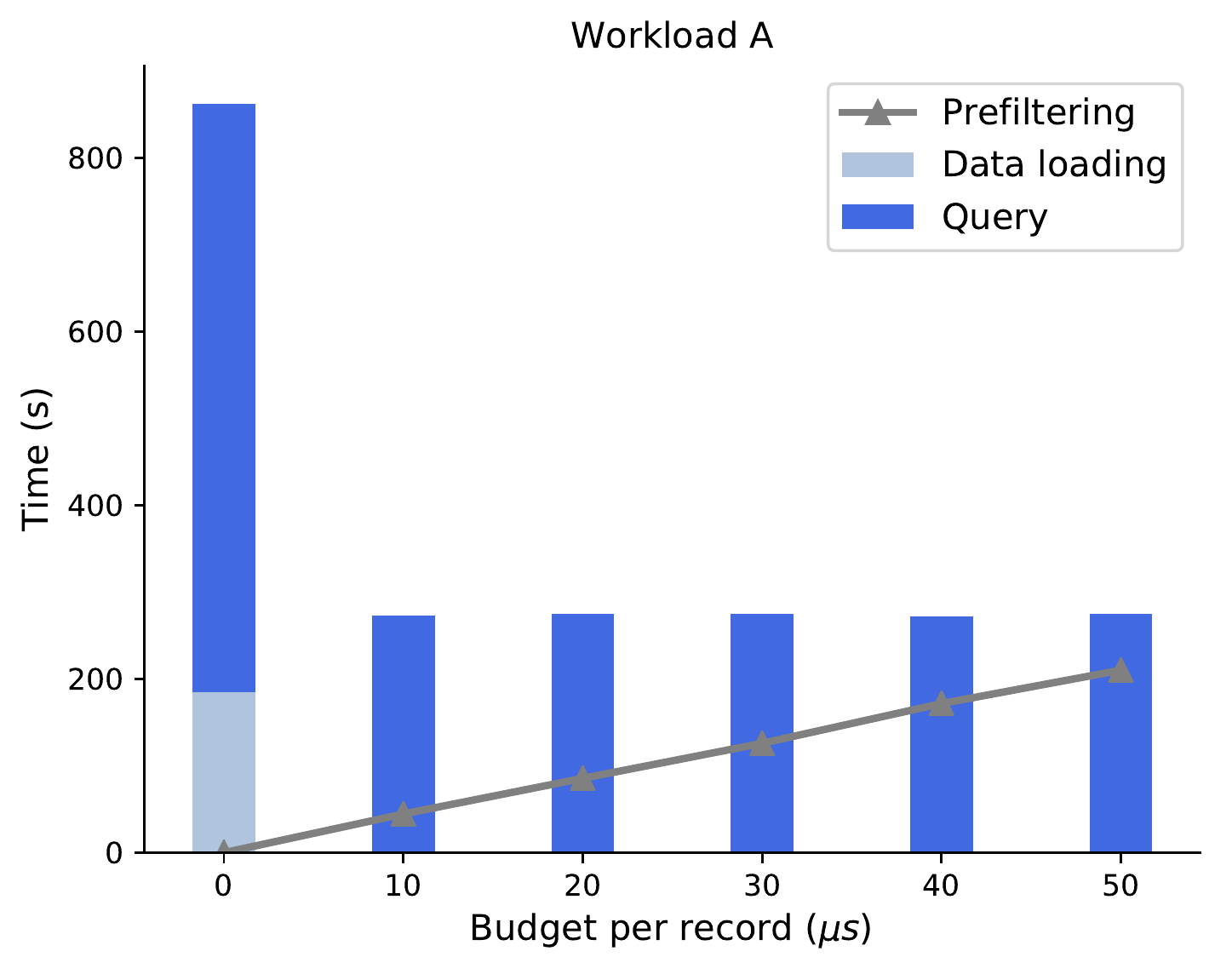}
       \includegraphics[height=40mm]{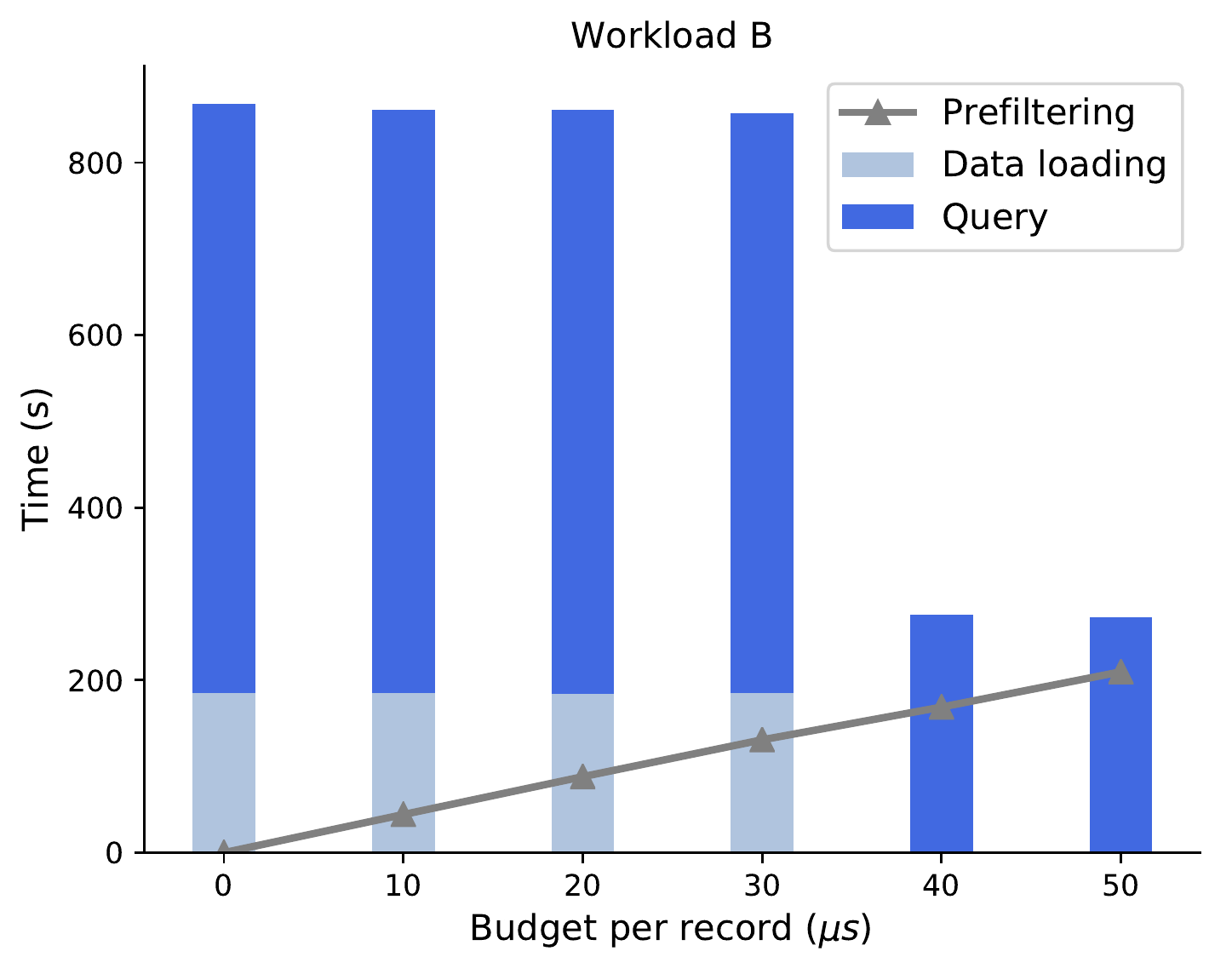}
       \includegraphics[height=40mm]{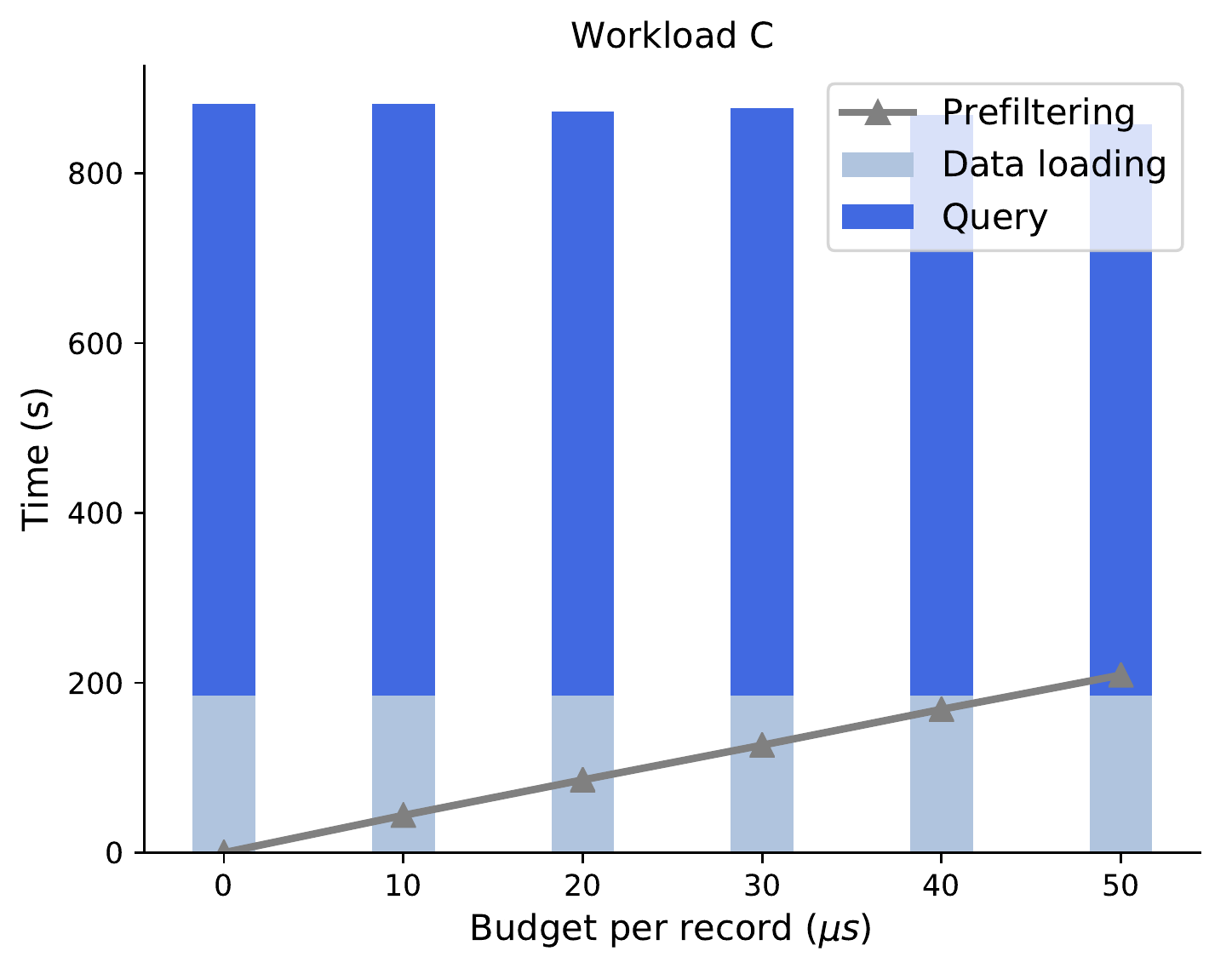}
      \caption{End-to-End Experiments of the 3 workloads on the Yelp Review Dataset}
      \label{fig:end2end-review}
\end{figure*}

\begin{figure*}[!t]
      \centering
       \includegraphics[height=40mm]{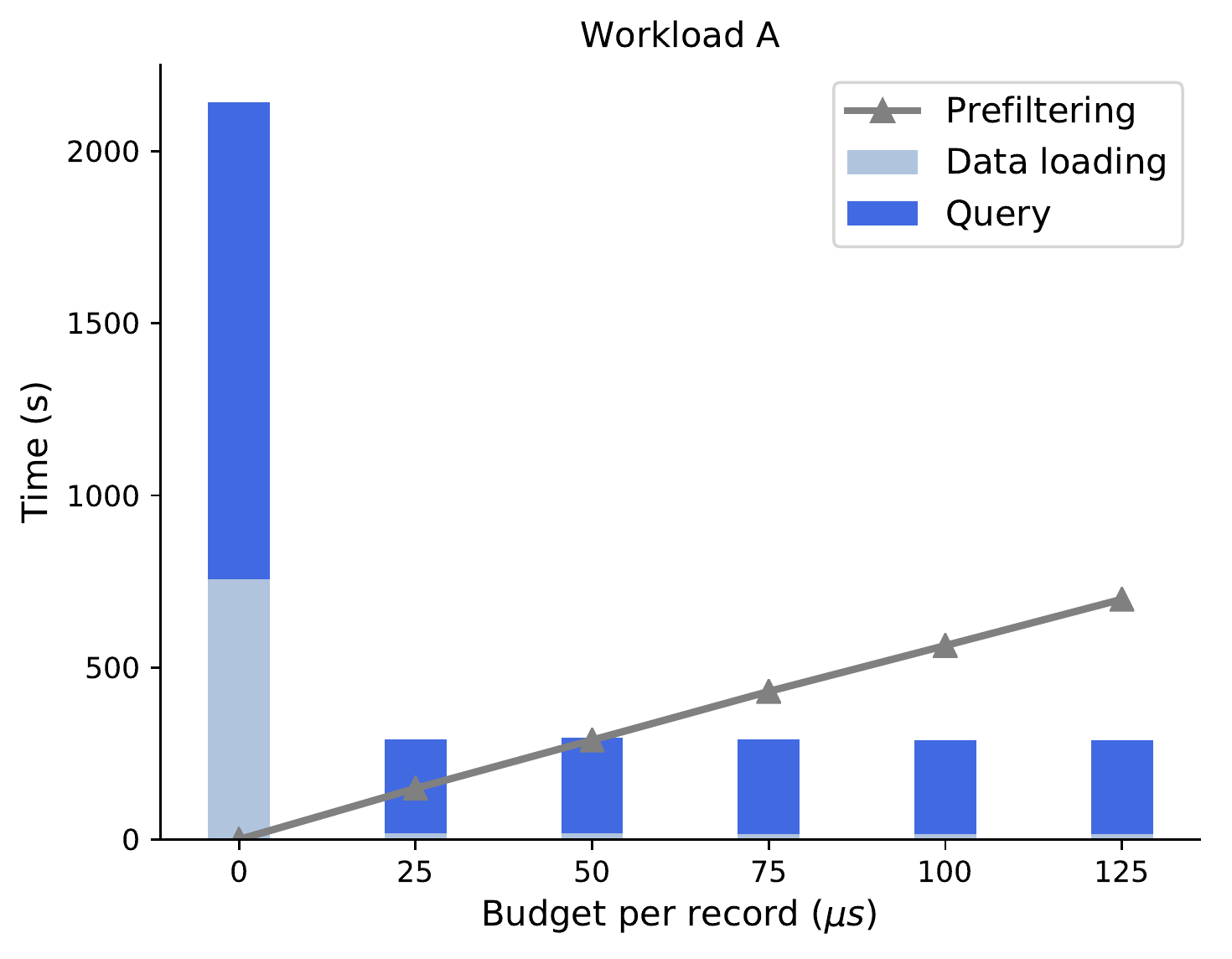}
       \includegraphics[height=40mm]{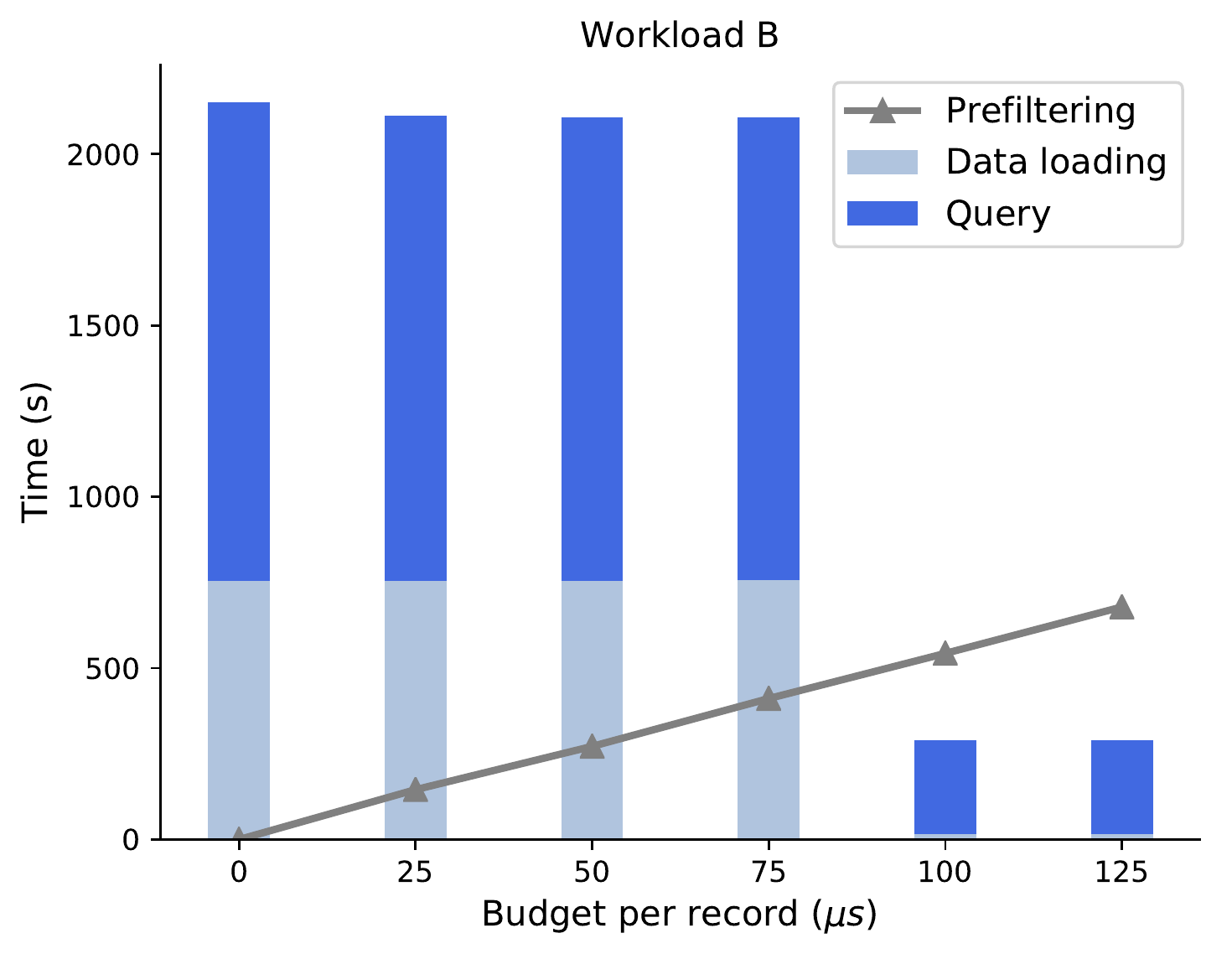}
       \includegraphics[height=40mm]{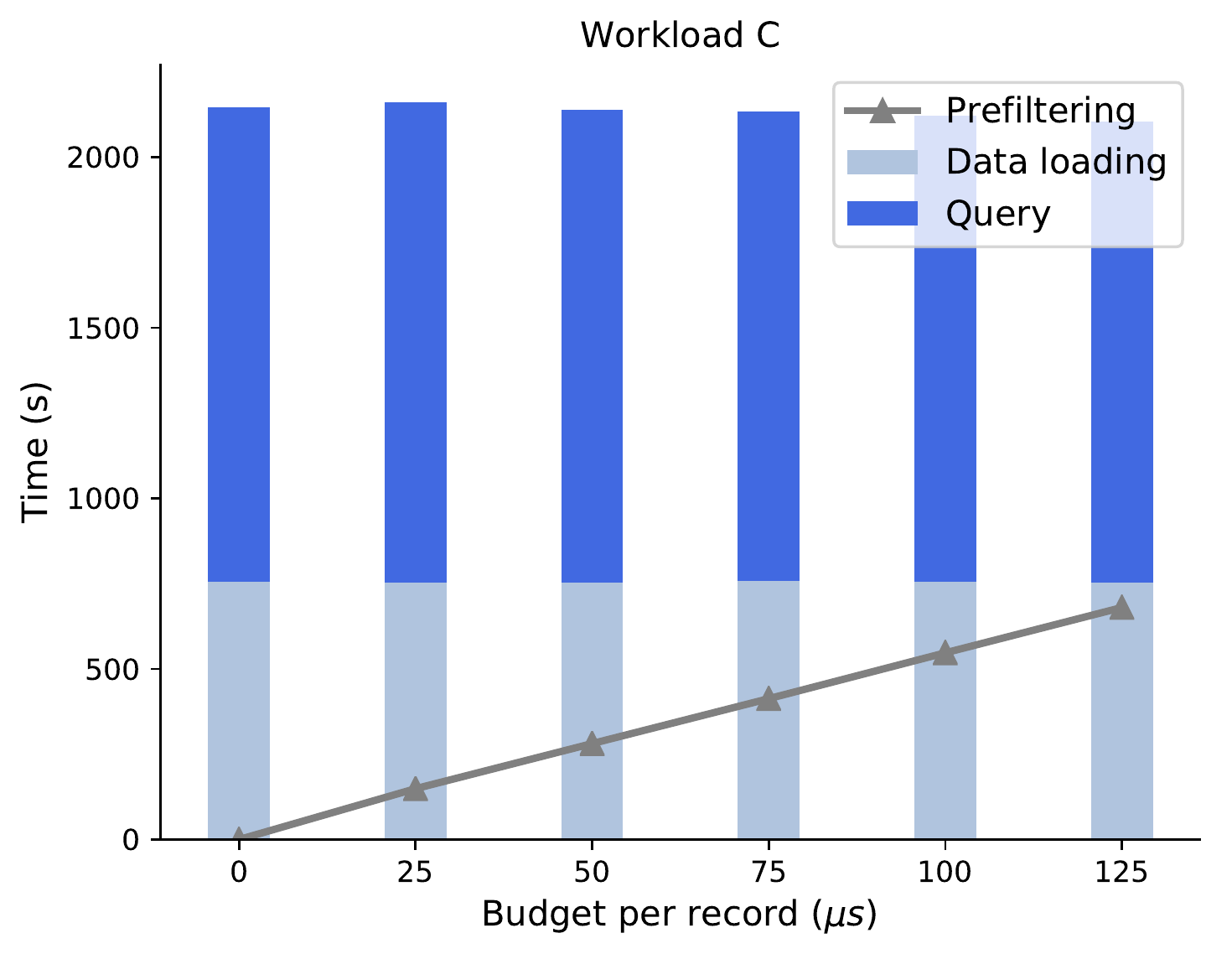}
      \caption{End-to-End Experiments of the 3 workloads on the YCSB Dataset}
      \label{fig:end2end-ycsb}
\end{figure*}

\subsection{Reducing data loading and query processing time with varied budgets}
\label{sec:end-to-end}

In this experiment, we generate three query workloads for each dataset to test the data loading time and query processing time of \sys under varied computation budgets. 
Each workload includes 200 queries and has a different distribution of the predicate skewness. 
Table~\ref{tab:workload} summarizes the information about three workloads.
The \texttt{\#Predicates} column shows the summation of the number of predicates that are included in all queries. 
The column \texttt{Min/Max \#Predicates} shows the minimum and the maximum number of predicates of a query. 
The last column shows the distribution of choosing the predicates. 
Here, workload A is highly skewed with a high predicate overlap, which represents the `easy' case where \sys can leverage predicate overlap and skewness to achieve the most benefit. 
On the other end, workload C has a low predicate overlap and the predicates are uniformly distributed, i.e. not skewed. 
Therefore, workload C represents the `challenging' case where \sys can be less beneficial. 
Finally, workload B is a middle ground of workload A and C, 
that is, it is less skewed than workload A. 
Note that we use Numpy to generate the Zipfian distribution and 
the smaller skewness parameter means higher skewness in its implementation 
(i.e. Table~\ref{tab:workload} shows that 
skewness parameters are 1.5 and 2 for workload A and B respectively.)

We first calibrate the cost model as shown in Section~\ref {sec:sensitivity_costmodel} for the current hardware configuration.
The cost model estimates the number of $\mu s$ of evaluating each predicate on a JSON object. 
For each dataset, we vary the computation budget and show how \sys performs given the same budget on different workloads. 
Our baseline in these experiments is the one with zero budget (i.e. no optimization is applied). 
The experiment results are shown in Fig.~\ref{fig:end2end-win}, Fig.~\ref{fig:end2end-review} and Fig~\ref{fig:end2end-ycsb} for the Windows System Log dataset, the Yelp Review dataset and the YCSB dataset respectively.

We observe the following trends in the experiment results. 
We see that the performance of the partial loading (denoted as \texttt{Data loading}) varies for different workloads.
For the `easy' workload A, partial loading is used even if the administrator uses a very small budget on clients. 
As expected, highly overlapped predicates and skewed distribution of predicates is beneficial for the optimization.
For workload B, a small budget limits the number of predicates that can be pushed down. 
In such a case, there may not be an opportunity for partial loading (i.e. there is no tuple that is invalid for all queries). 
As we increase the budget, more predicates are pushed down and the server can now avoid loading irrelevant JSON objects by utilizing the bit vectors.
For the workload C, the server does not employ partial data loading due to low predicates overlapping and low skewness. 
Among all the workloads for all datasets, \sys can accelerate the data loading process by 21x, 
with the budget of 1 $\mu$s.

In addition to partial data loading, our experiments show the results of the query processing time (denoted as \texttt{Query}). 
This is the total time to run the full workload of 200 queries.
We see that as we have a larger computation budget, we can push down more predicates to the client-side
and we can skip more tuples to reduce the query processing time. 
Specifically, \sys can accelerate query processing by 23x for the budget 1 $\mu$s.
Finally, we show the total time of evaluating the predicates on the client-side (denoted as \texttt{prefiltering}). 
We see that the prefiltering time is increased as we have a larger budget. 
 
\begin{figure}[t]
    \centering
      \includegraphics[width=0.7\linewidth]{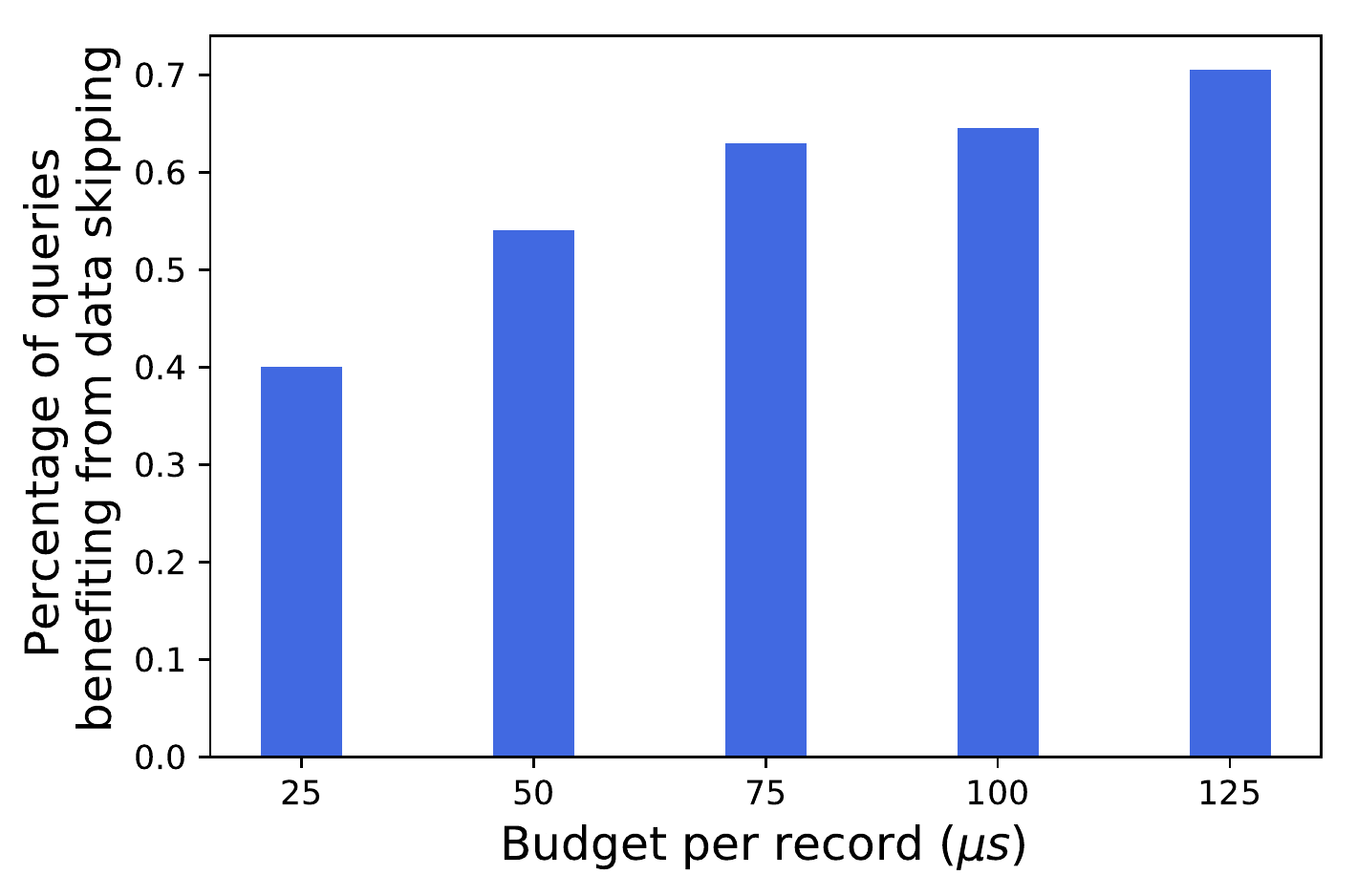}
      \caption{The percentage of queries that have less query processing time 
      due to data skipping for the workload C of YCSB dataset.}
      \label{fig:diff}
\end{figure}

We now further investigate the performance of \sys on the `challenging' workload using the YCSB dataset (i.e. Workload C). 
The aggregated result in the rightmost plot of Fig.~\ref{fig:end2end-ycsb} does not suggest a noticeable improvement. However, when we look at each query individually, we find that the data skipping technique can reduce the query processing time for several of the queries. 
To show this, we report the fraction of queries that have lower query processing time due to data skipping with varied computation budgets. 
The results are shown in Fig.~\ref{fig:diff}. 
We find that there are between 37\% and 68\% queries that benefit from the data skipping technique. 



\begin{figure}[!t]
    \centering
    \includegraphics[width=0.7\linewidth]{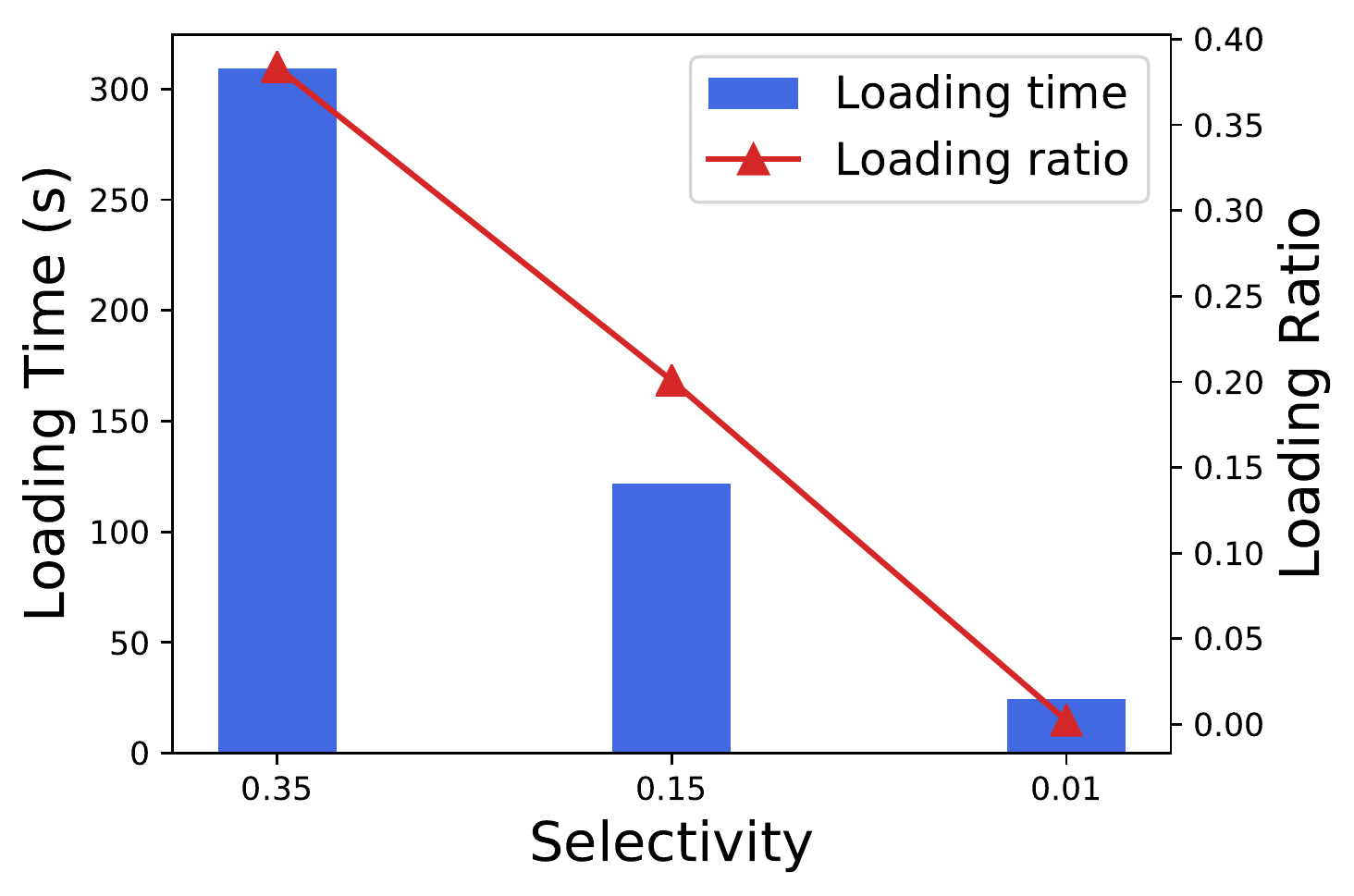}
      \caption{We vary the predicate selectivity and evaluate the data loading time on the Windows System Log dataset.}
      \label{fig:micro-sel-dl}
\end{figure}

\begin{figure}[!t]
    \centering
    \includegraphics[width=0.7\linewidth]{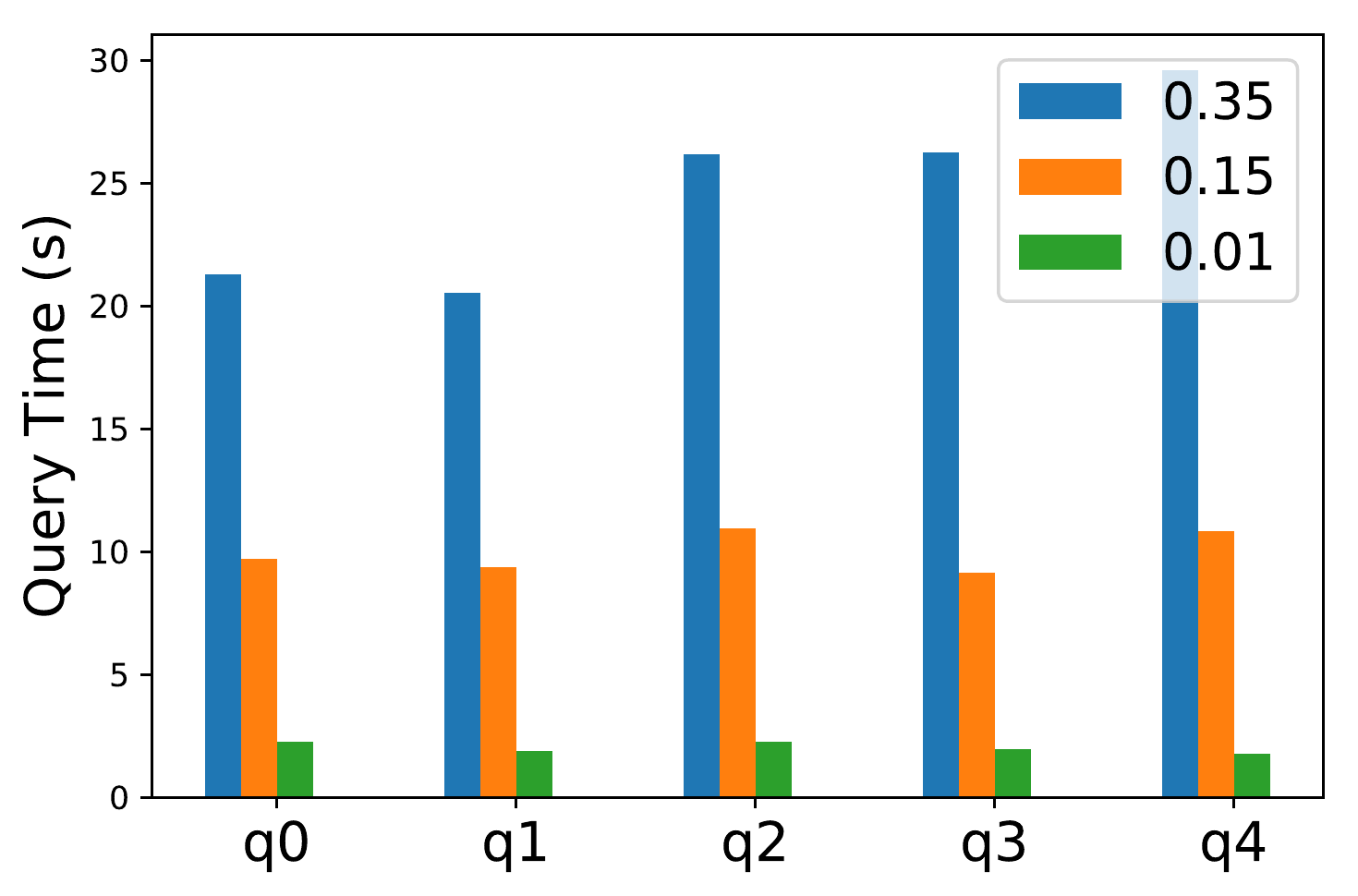}
      \caption{We vary the predicate selectivity and evaluate the query execution time on the Windows System Log dataset.}
      \label{fig:micro-sel-query}
\end{figure}

\subsection{Performance impact of predicate selectivity, overlap, and skewness}
\label{sec:micro}
In our micro-benchmarks, we break down the end-to-end experiments 
for a better understanding of how different characteristics of a workload affect the performance of \sys. 
For all the micro-benchmarks, we test the Windows System Log dataset.

\subsubsection{Sensitivity to predicate selectivity}
In this experiment, we evaluate how the selectivity of a workload affects the data loading time and the query execution time.
In the Windows System Log dataset, 
we generate predicates of different selectivities (0.01, 0.15, 0.35) by using attributes whose frequencies roughly represent the corresponding selectivity. 
We then construct 3 workloads, where each workload contains 5 queries and each query contains 3 conjunctive predicates. 
The queries in the high selectivity workload are of high selectivity (0.01) and vice versa for the medium (0.15) and low selectivity (0.35) workloads.


We push down 2 predicates to the client for each workload and make sure partial loading is enabled.
The results are shown in Fig.~\ref{fig:micro-sel-dl} and Fig.~\ref{fig:micro-sel-query} for data loading time and query execution time respectively. 
We note that the \texttt{loading ratio} is the ratio between the number of loaded objects and the number of all objects in the dataset.
We see the loading ratio changes due to predicate selectivity. 
This is because highly selective predicates will enable the server to load fewer JSON objects 
so the loading ratio will be relatively low, thus reducing loading time.
Fig.~\ref{fig:micro-sel-query} shows the query execution time for each query.
We see that as we decrease the selectivity (i.e. from 0.35 to 0.01), more tuples are skipped 
and the query processing time is reduced.

\begin{figure}[!t]
    \centering
    \includegraphics[width=0.7\linewidth]{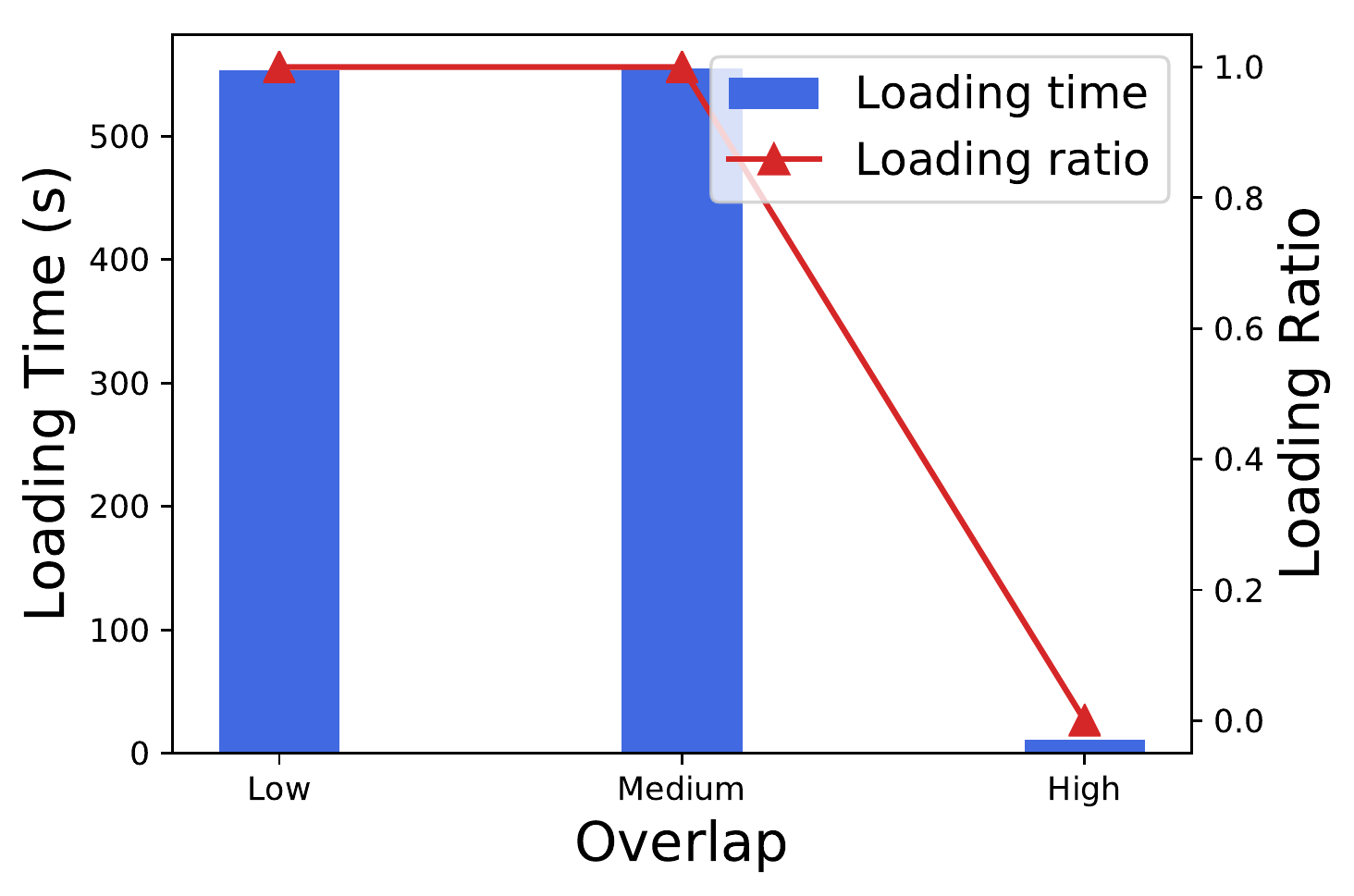}
      \caption{We vary the predicate overlapping and evaluate the data loading time on the Windows System Log dataset.}
      \label{fig:micro-ol-dl}
\end{figure}

\begin{figure}[!t]
    \centering
    \includegraphics[width=0.7\linewidth]{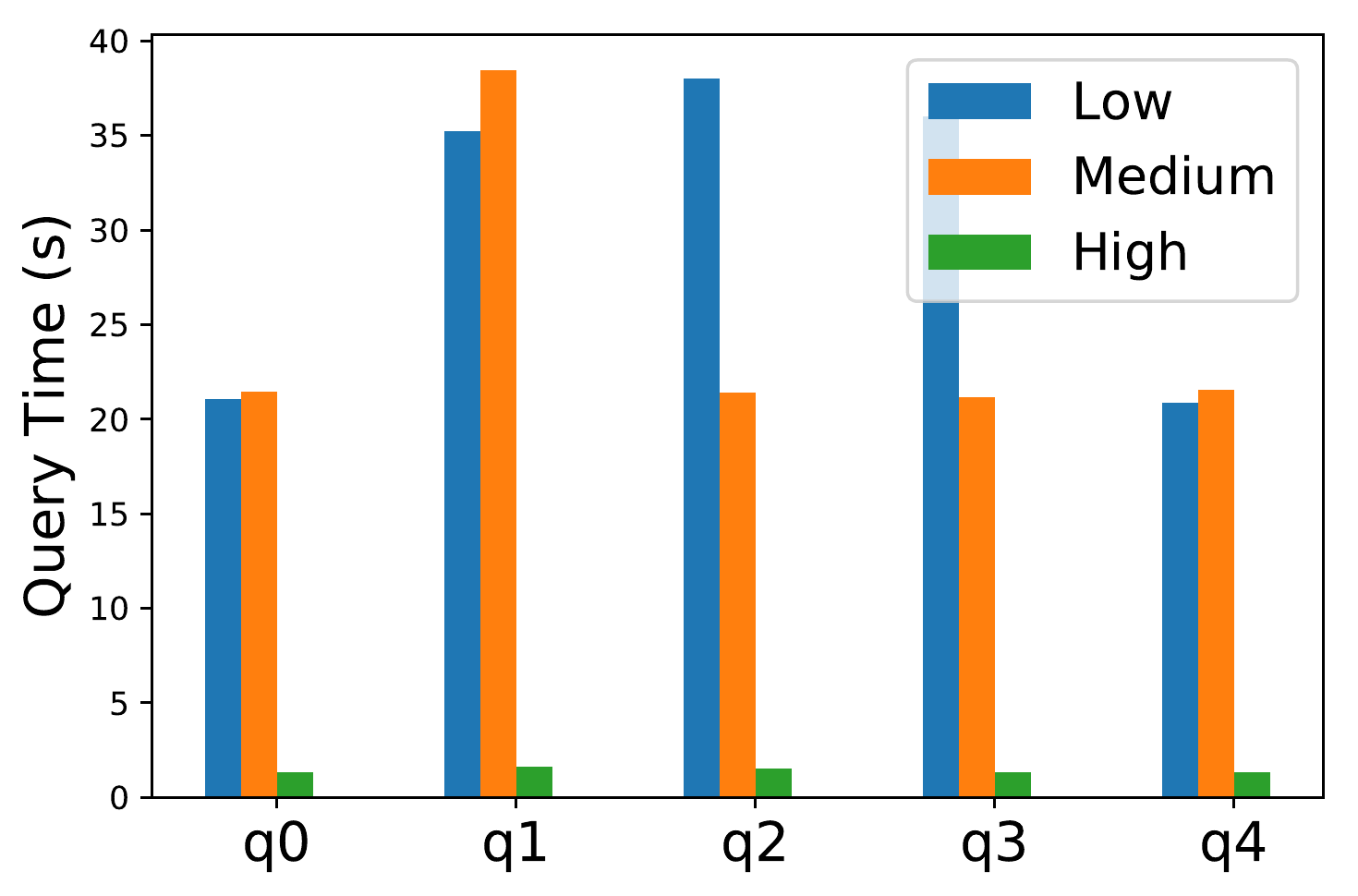}
      \caption{We vary the predicate overlapping and evaluate the query execution time on the Windows System Log dataset.}
      \label{fig:micro-ol-query}
\end{figure}

\subsubsection{Sensitivity to predicate overlap}

We next study how the predicate overlap impacts the data loading time and query processing time. 
We generate 3 workloads $L_{ol}$, $M_{ol}$, and $H_{ol}$. 
Each workload contains 5 queries and each query in workload $L_{ol}$, $M_{ol}$ and $H_{ol}$ includes 1, 2 and 4 conjunctive predicates respectively.
The predicates are distributed uniformly for all workloads and we push down two predicates for each workload.
In this setting, the workload $L_{ol}$ and $M_{ol}$ have low and medium predicate overlap respectively, 
and the workload $H_{ol}$ has a high predicate overlap.

The results for data loading time are shown in Fig.~\ref{fig:micro-ol-dl}. 
For the workload $L_{ol}$ and $M_{ol}$, the numbers of predicates pushed down are not large enough to enable partial loading. 
In practice, for a workload in which queries almost share no predicates, 
the administrator needs to set a large budget so that enough predicates can be pushed down to cover all the queries.
On the other hand, we can observe a drastic drop in loading time for workload $H_{ol}$. 
This is because queries in the workload $H_{ol}$ include more conjunctive predicates and the pushed-down predicates cover all the queries in the workload $H_{ol}$. 
Therefore, the server can employ partial data loading to reduce the data loading time. 

The results for query processing time are shown in Fig.~\ref{fig:micro-ol-query}. 
We see that even if we need to load all objects for both workloads $L_{ol}$ and $M_{ol}$, 
the workload with a higher predicate overlap (i.e. $M_{ol}$) reduces more query processing time.
This is because for $M_{ol}$ there are more queries (i.e. q2 and q3) 
that include the pushed-down predicates compared to workload $L_{ol}$. 
The workload $H_{ol}$ represents the best-case scenario in which we not only significantly reduce the data loading time due to partial loading, 
but also reduce the query processing time via data skipping.


\begin{figure}[!t]
    \centering
    \includegraphics[width=0.7\linewidth]{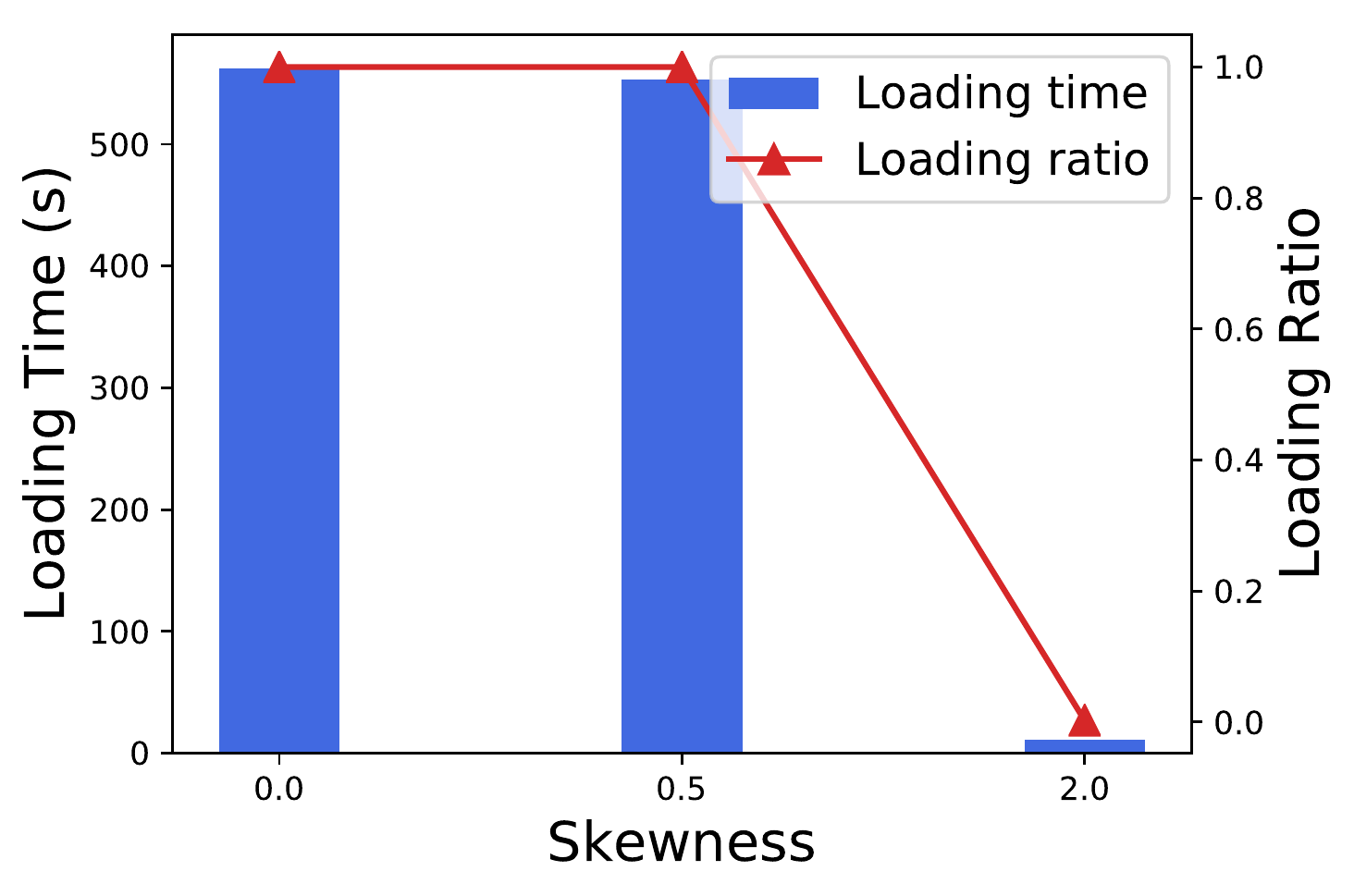}
      \caption{We vary the predicate skewness and evaluate the data loading time on the Windows System Log dataset.}
      \label{fig:micro-sk-dl}
\end{figure}

\begin{figure}[!t]
    \centering
    \includegraphics[width=0.7\linewidth]{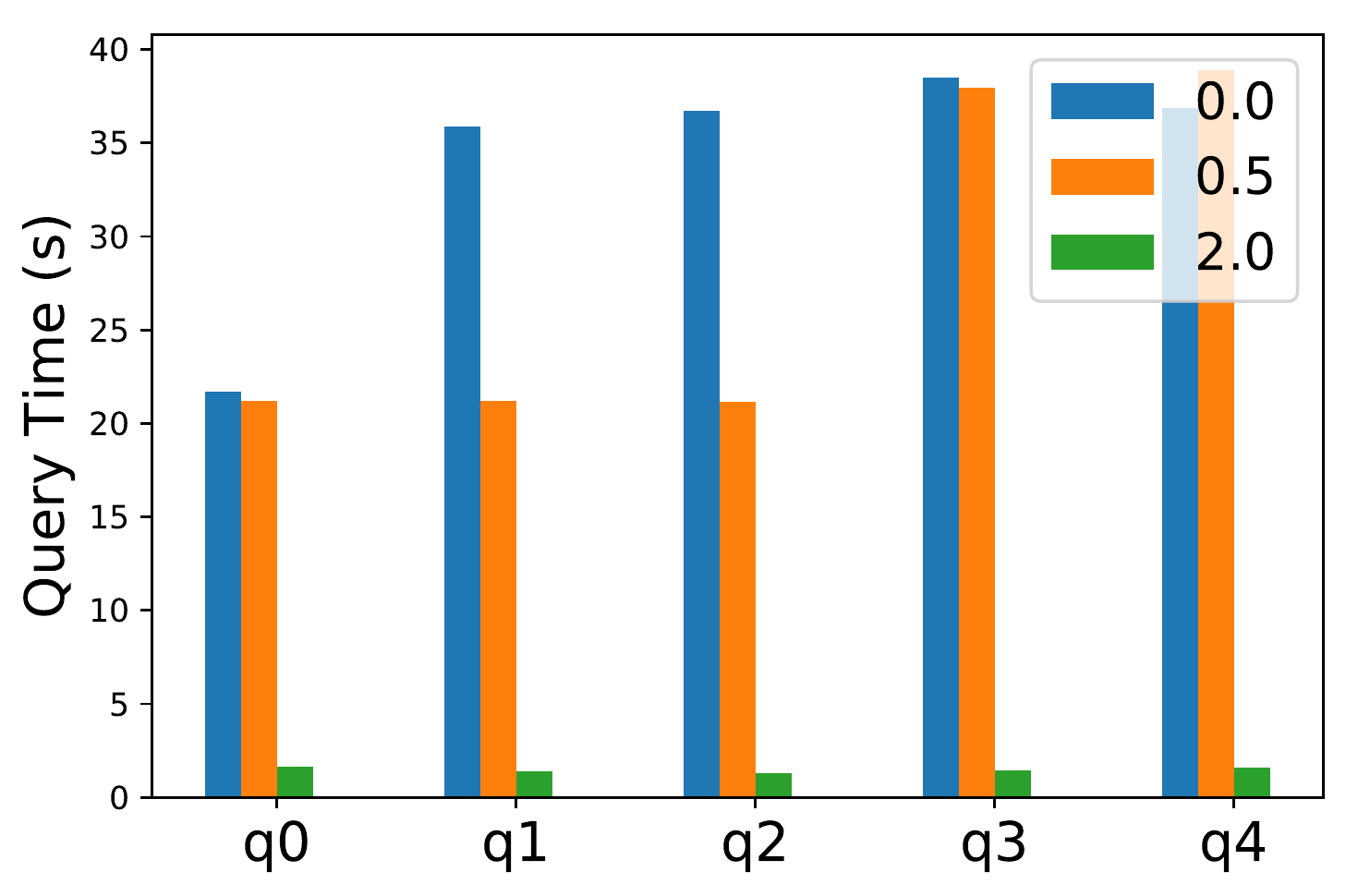}
      \caption{We vary the predicate skewness and evaluate the query execution time on the Windows System Log dataset.}
      \label{fig:micro-sk-query}
\end{figure}

\subsubsection{Sensitivity to predicate skewness}

A highly skewed distribution of predicates can be beneficial for \sys 
because a small number of conjunctive predicates are included in all queries. 
In the most extreme scenario, one predicate that is pushed down is included in all queries. 
In this case, the server can directly skip all tuples that fail to satisfy the predicate.

We can calculate the skewness factor with the following formula:
\begin{equation*}
\begin{aligned}
    \frac{\sum_{i=1}^N(X_i-\bar{X})^3}{(N-1)\sigma^3}
\end{aligned}
\end{equation*}
where $N$ is the number of distinct predicates, 
$X_i$ denotes number of queries including the predicate $i$, 
$\bar{X}$ denotes the expected value of $X_i$, 
$\sigma$ indicates the standard deviation of predicate distribution, i.e., $\sigma = \sqrt{\frac{\sum_{i=1}^N(X_i-\bar{X})^2}{N}}$.

To evaluate how skewness impacts the performance of \sys, we construct three workloads of 5 queries, where each query includes 2 predicates.
We set the skewness factors for workload $L_{sk}$, $M_{sk}$ and $H_{sk}$ to 0.0, 0.5 and 2.0 respectively 
and we only push one predicate down to the client.

The results for data loading time and query processing time are shown in Fig.~\ref{fig:micro-sk-dl} and Fig.~\ref{fig:micro-sk-query} respectively. 
For different skewness levels, the different number of queries include the predicates that are pushed down:
\begin{itemize}
    \item for workload $L_{sk}$, q0 includes the pushed down predicate; therefore, 
    q0 is the only query benefiting from data skipping;
    \item for workload $M_{sk}$, q0, q1, and q2 are covered and this leads to a drop in query processing time for workload $M_{sk}$ as shown in Fig.~\ref{fig:micro-sk-query};
    \item for workload $H_{sk}$, all queries are covered and the server enables partial loading to reduce the data loading time as shown in Fig.~\ref{fig:micro-sk-dl}.
\end{itemize}

\begin{table}[t]
\caption{We calibrate the cost model under three different hardware environments and report R-squared values.}
\centering
\begin{tabular}{@{}ccc@{}}
\toprule
\textbf{Platform} & \textbf{Hardware}                                                                            & \textbf{R-squared} \\ \midrule
Local Server      & \begin{tabular}[c]{@{}c@{}}2-core Intel Core i7-5557U @ 3.10 GHz\\ RAM: 16 GB\end{tabular}    & 0.897              \\ \midrule
Alibaba Cloud     & \begin{tabular}[c]{@{}c@{}}4 vCPU-core Intel Xeon @ 2.5 GHz\\ RAM: 8 GB\end{tabular}         & 0.666              \\ \midrule
PKU Weiming       & \begin{tabular}[c]{@{}c@{}}32-core Intel Xeon Gold 6240 @ 2.6 GHz\\ RAM: 192 GB\end{tabular} & 0.978              \\ \bottomrule
\end{tabular}
\label{tab:cost-model}
\end{table}

\subsection{Robustness of the cost model}
\label{sec:sensitivity_costmodel}
In this experiment, we demonstrate the robustness of our cost model discussed in Sec. \ref{sec:costmodel} using different hardware configurations.
We calibrate the cost model under three different hardware environments as shown in Table \ref{tab:cost-model}.
We randomly choose 100 predicates respectively from three datasets and select a sample with a size of 5 GB for each dataset.
The client evaluates the predicates and records the time cost and selectivity for each predicate.
Then we conduct multivariate linear regression on the results and compute the coefficients for the specific hardware environment.

We construct the cost model on three different hardware platforms: 
an ``on-premise'' server, Alibaba Cloud ECS (Elastic Compute Service) 
and Weiming Teaching Cluster of Peking University.
Specific details of hardware environments are listed in Table~\ref{tab:cost-model}.

We use R-squared, a common statistical measure in linear regression to show how well the data fits the regression model. 
R-squared can be calculated as follows:
\begin{equation*}
\begin{aligned}
    R^2 = 1 - \frac{\sum_{i}(\hat{y_i} -y_i)^2}{\sum_{i}(\hat{y_i} -\bar{y})^2}
\end{aligned}
\end{equation*}

We see that the results on the local server and PKU Weiming Cluster fit the model very well, 
while the cost model on Alibaba Cloud does not work as good as the other two configurations. 
We believe this is largely due to not running on bare-metal, but instead, an opaque hypervisor that can limit computation cycles or even migrate the virtual machine while running.

\section{Conclusion}
\label{sec:conc}
Analytics databases often centralize data collected. Data loading (i.e., parsing, validating, and storing the client data) is an under-appreciated bottleneck on a database server.
There are numerous opportunities to offload parts of the loading process to client devices to reduce the perceived end-to-end latency in query processing.
CIAO evaluates query predicates on client-devices without fully parsing the data on the clients.
It leverages string pattern-matching primitives to directly query JSON records on the client devices, and allows for data skipping and partial data loading.
Our experimental results show that the system substantially accelerates data loading by up to 21x and query execution by up to 23x and improves end-to-end performance by up to 19x within a budget of 1.0 microseconds latency per record on clients.

\bibliographystyle{IEEEtran}
\bibliography{reference}
\end{document}